\documentclass[aoas,MSNbibl,nameyear,noautosecdot,dvips]{arximspdf}
\usepackage{dcolumn}
\usepackage{graphicx}

\doi{10.1214/11-AOAS508}
\volume{6}
\issue{1}
\pubyear{2012}
\firstpage{83}
\lastpage{105}

\makeatletter

\def\bsuffix #1{#1}

\newcolumntype{d}[1]{D{.}{.}{#1}}

\newtheorem{proposition}{Proposition}

\makeatother

\begin{document}
\begin{frontmatter}

\title{An exact adaptive test with superior design sensitivity in
an observational study of~treatments for ovarian cancer}
\runtitle{Adaptive tests and design sensitivity}

\begin{aug}
\author[A]{\fnms{Paul R.} \snm{Rosenbaum}\corref{}\thanksref{t1}\ead[label=e1]{rosenbaum@wharton.upenn.edu}}
\runauthor{P. R. Rosenbaum}
\affiliation{University of Pennsylvania}
\address[A]{Department of Statistics\\
The Wharton School\\
University of Pennsylvania\\
473 Jon M. Huntsman Hall\\
3730 Walnut Street\\
Philadelphia, Pennsylvania 19104-6340\\
USA\\
\printead{e1}} 
\end{aug}

\thankstext{t1}{Supported by a grant from the NSF.}

\received{\smonth{5} \syear{2011}}
\revised{\smonth{8} \syear{2011}}

%
\vspace*{6pt}
\begin{abstract}
A sensitivity analysis in an observational study determines the
magnitude of bias from nonrandom treatment assignment that would need
to be present to alter the qualitative conclusions of a na\"{\i}ve
analysis that presumes all biases were removed by matching or by other
analytic adjustments. The power of a sensitivity analysis and the
design sensitivity anticipate the outcome of a sensitivity analysis
under an assumed model for the generation of the data. It is known that
the power of a sensitivity analysis is affected by the choice of test
statistic, and, in particular, that a statistic with good Pitman
efficiency in a randomized experiment, such as Wilcoxon's signed rank
statistic, may have low power in a sensitivity analysis and low design
sensitivity when compared to other statistics. For instance, for an
additive treatment effect and errors that are Normal or logistic or
$t$-distributed with 3 degrees of freedom, Brown's combined quantile
average test has Pitman efficiency close to that of Wilcoxon's test but
has higher power in a sensitivity analysis, while a version of
Noether's test has poor Pitman efficiency in a randomized experiment
but much higher design sensitivity so it is vastly more powerful than
Wilcoxon's statistic in a sensitivity analysis if the sample size is
sufficiently large. A new exact distribution-free test is proposed that
rejects if either Brown's test or Noether's test rejects after
adjusting the two critical values so the overall level of the combined
test remains at $\alpha$, conventionally $\alpha=0.05$. In every
sampling situation, the design sensitivity of the adaptive test equals
the larger of the two design sensitivities of the component tests. The
adaptive test exhibits good power in sensitivity analyses
asymptotically and in simulations. In one sampling situation---Normal
errors and an additive effect that is three-quarters of the standard
deviation with 500 matched pairs---the power of Wilcoxon's test in a
sensitivity analysis was 2\% and the power of the adaptive test was
87\%. A study of treatments for ovarian cancer in the Medicare
population is discussed in detail.
\end{abstract}

%

\end{frontmatter}

\section{\texorpdfstring{Introduction: Motivation; example; outline.}{Introduction: Motivation; example; outline}}

\label{secIntro}

\subsection{\texorpdfstring{Are large observational studies less susceptible to unmeasured
biases\textup{?}}{Are large observational studies less susceptible to unmeasured
biases}}

\label{ssIntroAreLarge}

There is certainly a sense in which large observational studies are
more---not less---susceptible to unmeasured biases than smaller
studies. Biases
due to nonrandom treatment assignment generally do not become smaller
as the
sample size increases. These biases are due to the failure to control some
unmeasured covariate that would have been balanced by random assignment of
treatments. If a large observational study is analyzed na\"{\i}vely under
the assumption that adjustments for measured covariates have, in effect,
transformed the study into a randomized experiment, then as the sample size
increases even very small biases due to unmeasured covariates can seriously
distort the level of significance tests and the coverage of confidence
intervals; see Cochran (\citeyear{Coc}), Section 3.1.

Suppose, however, that the analysis takes explicit account of uncertainty
about unmeasured biases by performing a sensitivity analysis. Is a large
sample size of any assistance in this case? It is known that the degree of
sensitivity to unmeasured biases is affected by many aspects of the
design and
analysis of an observational study [Rosenbaum (\citeyear{Ros04}, \citeyear{Ros10N2})], but the relevant
decisions about design and analysis are often difficult to make without
guidance from empirical data. \citet{HelRosSma09}
found that sample
splitting---sacrificing a small portion, say, 10\%, of the sample to guide
design and analysis---could, in favorable circumstances, yield reduced
sensitivity to unmeasured biases by guiding the needed decisions. Sample
splitting has the advantage, emphasized by \citet{Cox75}, of permitting
reflection and judgement in light of data without invalidating the formal
properties of statistical procedures. However, some questions, such as the
thickness of the tails of distributions, are difficult to settle using
a~small
fraction of the sample, and may require guidance from the complete sample.
Here, an adaptive test is proposed that chooses between two tests with
different properties, and in one sense achieves the performance of the better
test in large samples; see Proposition~\ref{PropDS} in Section \ref
{ssAdaptiveDS}.
Although motivated by large sample calculations, the adaptive procedure
performs well in simulations in samples as small as 100 matched
pairs.

\subsection{\texorpdfstring{Example: Is more chemotherapy for ovarian cancer more effective\textup{?}}{Example: Is more chemotherapy for ovarian cancer more
effective}}

\label{secEG}

Following surgery to remove a visible tumor, the typical reason that
one cancer
patient receives more chemotherapy than another is that their cancers differ
in localization or recurrence. A straightforward comparison of patients
receiving more or less chemotherapy is likely to be biased by comparing sicker
patients to healthier ones. Is there a better comparison? Ovarian
cancer is
unusual in this regard, because there is a source of variation in the
intensity of chemotherapy that is not a reaction to the patient and her
illness. Chemotherapy for ovarian cancer may be provided by either a medical
oncologist who treats cancers of all kinds or by a gynecological oncologist
who treats cancers of the ovary, uterus and cervix. Medical oncologists
(MOs) and gynecological oncologists (GOs) differ in both training and
practice. In particular, GOs are gynecologists, and hence surgeons, perhaps
the best surgeons for gynecological cancers, and they often perform surgery
for ovarian \mbox{cancer}, whereas MOs are almost invariably not surgeons and
administer chemotherapy after someone else, perhaps a general surgeon, a
gynecologist or~GO, has performed surgery. Typically, an MO had a residency
in internal medicine followed by a 3-year fellowship in oncology emphasizing
the use of chemotherapy, whereas a GO had a residency in obstetrics and
gynecology followed by a fellowship in gynecologic oncology with attention
paid to surgical treatment of ovarian cancer. \citet{RosRosSil07N2}
hypothesized correctly that MOs would use chemotherapy more
intensively than
GOs, and they used this difference in intensity to ask whether more
chemotherapy is of benefit to the patient.

\begin{figure}

\includegraphics{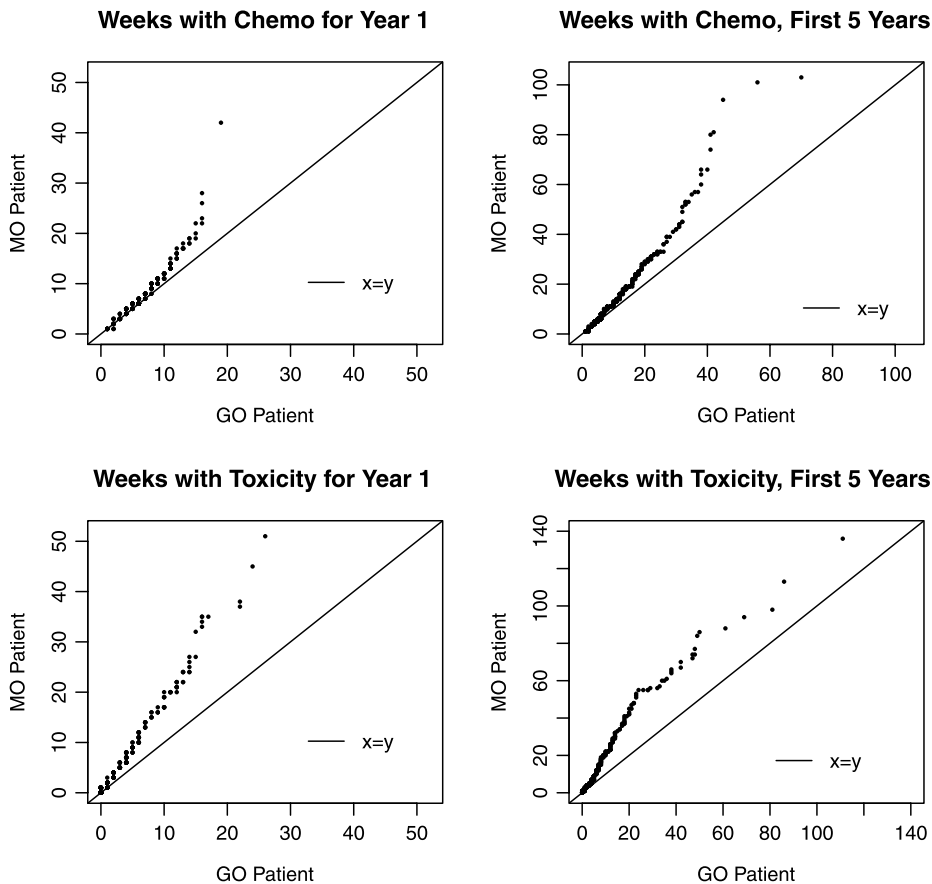}

\caption{Four quantile--quantile plots of weeks with either chemotherapy or
toxicity for $I=344$ pairs of an MO and a GO patient.
Quantile--quantile plots describe the marginal distributions, ignoring
who is paired with whom.}\label{fig1}
\end{figure}

Using data from Medicare and the Surveillance, Epidemiology and End
Results (SEER) program of the U.S. National Cancer Institute,
\citet{RosRosSil07N2} looked at patients with ovarian cancer
between 1991 and 2001 who received chemotherapy after appropriate
surgery; see their paper for details of the patient population. They
matched all $I=344$ such ovarian cancer patients treated by a
gynecologic oncologist to 344 ovarian cancer patients treated by a
medical oncologist. Using the matching algorithm of
\citet{RosRosSil07N1}, the matching controlled for 36 covariates,
including clinical stage, tumor grade, surgeon type, comorbid
conditions such as diabetes and congestive heart failure, age, race and
year of diagnosis [\citet{RosRosSil07N2}, Tables 2 and 3].
Importantly, the duration of follow-up was virtually identical in the
two groups. On average, during the five years after diagnosis, the
patients of medical oncologists received about four more weeks of
chemotherapy, with MO patients receiving on average 16.5 weeks of
chemotherapy and GO patients receiving 12.1 weeks. The upper portion of
Figure~\ref{fig1} is a pair of two quantile--quantile plots
[\citet{WilGna68}] of weeks of chemotherapy in the first year or
the first five years for the 344 GO patients and the 344 MO patients,
momentarily ignoring who is matched to whom. Because the points lie
above the line of equality, the distribution of chemotherapy weeks for
MO patients appears to be stochastically larger than the distribution
for GO patients. Survival was virtually identical with nearly identical
Kaplan--Meier survival curves that crossed repeatedly, and a median
survival of 2.98 years in the MO group and 3.04 years in the GO group
[\citet{RosRosSil07N2}, Figure 1 and Table~1]. Patients of medical
oncologists experienced more weeks with chemotherapy associated side
effects or toxicity, such as anemia, neutropenia, thrombocytopenia and
drug induced neuropathy, on average over five years, 16.2 weeks for MOs
and 8.9 weeks for GOs; see the bottom half of Figure 2. If Wilcoxon's
signed rank test is used to compare weeks with toxicity in matched
pairs, the $P$-values are less than $10^{-6}$ for both year one and the
first five years, but of course those $P$-values take no account of
possible biases in this nonrandomized comparison. In brief, greater
intensity of chemotherapy was not associated with longer survival, but
it was associated with more frequent side effects.

The study generated some discussion, in particular, an editorial, five letters
discussing either the study or the editorial, and two rejoinders, one
from the
authors of the paper and one from the author of the editorial, or 11
pages of
published discussion of a 7 page paper. Happily, matching for 36 measured
covariates was convincing in the very limited sense such adjustments
can be
convincing: none of the discussion expressed continued concern about
these 36
measured covariates, which include many of the key covariates for ovarian
cancer. Virtually all of the discussion concerned possible unmeasured
biases, possible ways the MO and GO groups may differ besides the 36 measured
covariates. The editorial by an MO mentioned the magnitude of
``residual tumor'' not removed by surgery, a
covariate not recorded in SEER, and suggested the possibility that GOs were
less prone to notice toxicity, whereas the first letter by two GOs
characterized these comments as ``spinning a~tale.'' The particulars of
the discussion had strengths and
weaknesses, but, in an abstract sense, a concern about possible unmeasured
biases is reasonable in most if not all observational studies, and in that
sense the discussion was constructively focused on the central issue. A
disappointing feature of the 11 pages of discussion was that it contained
little in the way of data, quantitative analysis or evidence, although there
was a little data in one rejoinder. A sensitivity analysis in an
observational study is an attempt to return to the data and to quantitative
analysis when discussing the possible impact of unmeasured biases.

How large would the departure from random assignment have to be to
alter the conclusions? The answer is determined by a sensitivity
analysis. The degree of sensitivity to unmeasured biases in this study
is noticeably affected by the choice of test statistic; see Section
\ref{secExampleAnalysis}. Theoretical considerations suggest that
certain statistics, for instance, Wilcoxon's statistic, tend to
exaggerate the degree of sensitivity to unmeasured biases, at least for
additive treatment effects with symmetric errors
[\citet{Ros10N1}], so perhaps certain methods may be excluded on
purely theoretical grounds. On the other hand, many issues affect the
degree of sensitivity to bias reported by different test statistics
[\citet{Ros10N2}, Part III], and some of these issues are
difficult to evaluate prior to looking at the data. Here, an exact,
adaptive test is proposed that chooses, after the fact, the less
sensitive of two analyses, exactly correcting the level of the test for
the use of two analyses. Is adapting the test statistic to the data at
hand of value in sensitivity analyses?

\subsection{\texorpdfstring{Outline: Review; an exact adaptive test; design sensitivity;
power.}{Outline: Review; an exact adaptive test; design sensitivity;
power}}

Section~\ref{secReview} is a review of existing background material and
notation, including randomization inference in experiments in Section
\ref{ssReviewRand}, sensitivity analysis in observational studies in
Section \ref{ssReviewSen}, and the power of a sensitivity analysis and
the design sensitivity in Section \ref{ssReviewDS}; there is little new
material in the review in Section \ref{secReview}. With notation and
background established, Section \ref{ssWhyImportant} discusses why
adaptation is important in this context. The new adaptive test is
discussed in Section \ref{secAdaptive}, its exact null distribution in
Sections~\mbox{\ref{ssExactNulol}--\ref{ssNumericalNull}}, its nonnull
asymptotic properties in Section~\ref{ssAdaptiveDS}, and its finite
sample power obtained by simulation in Section~\ref{ssSimulation}. In
particular, Proposition~\ref{PropDS} of Section~\ref{ssAdaptiveDS}
shows that in each sampling situation, the design sensitivity of the
adaptive procedure is equal to the maximum of the design sensitivities
of the two nonadaptive procedures from which it is built. The
simulation suggests that the asymptotic properties begin to take effect
in samples of modest size. In Section \ref{secExampleAnalysis} the
methods are applied to the example in Section~\ref{secEG} from
\citet{RosRosSil07N2}. The discussion in Section \ref{secDisc}
considers related alternative methods in Section \ref{ssDiscVariations}
and returns in Section \ref {ssDiscAreLarge} to the question raised in
Section \ref{ssIntroAreLarge}.

\section{\texorpdfstring{Notation and review: Randomization; sensitivity analysis; design
sensitivity.}{Notation and review: Randomization; sensitivity analysis; design
sensitivity}}

\label{secReview}

\subsection{\texorpdfstring{Inference about treatment effects in a randomized experiment.}{Inference about treatment effects in a randomized experiment}}

\label{ssReviewRand}

There are $I$ matched pairs, $i=1,\ldots,I$, of two subjects, $j=1,2$, one
treated with $Z_{ij}=1$, the other control with $Z_{ij}=0$, so $Z_{i1}%
+Z_{i2}=1$ for each $i$. The subjects were matched for an observed
covariate, $\mathbf{x}_{ij}$, so $\mathbf{x}_{i1}=\mathbf{x}_{i2}$ for
each~$i$, but they may differ in terms of an unobserved covariate
$u_{ij}$, so possibly $u_{i1}\neq u_{i2}$. Subject $ij$ has two
potential responses, namely, $r_{Tij}$ if $ij$ is assigned to treatment
with \mbox{$Z_{ij}=1$} and $r_{Cij}$ if $ij$ is assigned to control with
$Z_{ij}=0$, so the response observed from $ij$ is $R_{ij}=Z_{ij}
r_{Tij}+( 1-Z_{ij}) r_{Cij}$ and the effect of the treatment
$r_{Tij}-r_{Cij}$ on subject $ij$ is not observed for any subject; see
Neyman (\citeyear{Neyman23}), \citet{Wel37}, \citet{Rub74},
\citet{Rei00} and \citet{Gad01}. Fisher's (\citeyear{Fis35})
sharp null hypothesis $H_{0}$ of no treatment effect asserts
$H_{0}\dvtx r_{Tij}=r_{Cij}$, for $i=1,\ldots,I$, $j=1,2$, whereas the
hypothesis~$H_{\tau}$ of an additive constant treatment effect $\tau$
asserts $H_{\tau}\dvtx r_{Tij}=r_{Cij}+\tau $ for all $ij$.

Write $\mathcal{F}=\{ ( r_{Tij},r_{Cij},\mathbf{x}_{ij}%
,u_{ij}) , i=1,\ldots,I, j=1,2\} $ for the potential responses
and covariates and write $\mathcal{Z}$ for the event that $(
Z_{i1}+Z_{i2}=1, i=1,\ldots,I) $. In a randomized paired experiment,
one subject in each pair $i$ is picked at random to receive the
treatment, so
$\Pr( Z_{ij}=1\vert\mathcal{F},\mathcal{Z})
=\frac{1}{2}$ for each~$ij$, with independent assignments in distinct pairs.

Within pair $i$, the treated-minus-control difference $Y_{i}$ in observed
responses is%
\begin{eqnarray*}
Y_{i} & = & ( Z_{i1}-Z_{i2}) ( R_{i1}-R_{i2})
=Z_{i1}( r_{Ti1}-r_{Ci2}) +Z_{i2}( r_{Ti2}-r_{Ci1})
\\
& = & \tau+\varepsilon_{i}\qquad\mbox{with }\varepsilon_{i}=( Z_{i1}-Z_{i2})
( r_{Ci1}-r_{Ci2}) \qquad\mbox{if }H_{\tau}\mbox{ is true.}%
\end{eqnarray*}
Given $\mathcal{F},\mathcal{Z}$, the quantity $r_{Ci1}-r_{Ci2}$ is
fixed, and
in a randomized experiment $\varepsilon_{i}=\pm\vert r_{Ci1}-r_{Ci2}%
\vert$ with equal probabilities $\frac{1}{2}$, so if $H_{\tau}$ is true,
then $Y_{i}$ is symmetric about $\tau$.

Ties among the $Y_{i}$'s are not a problem, but the development is
simpler if
ties of all kinds are assumed absent. In particular, when testing
$H_{\tau}%
$, the $\vert Y_{i}-\tau\vert$ are assumed to be untied, and the
$Y_{i}-\tau$ are assumed to not equal zero. Minor adjustments in
Section \ref{ssTies} eliminate these restrictions.

When testing $H_{\tau}$, let $q_{i}$ be the rank of $\vert Y_{i}%
-\tau\vert$ and let $S_{i}=1$ if $Y_{i}-\tau>0$ or $S_{i}=0$ if
$Y_{i}-\tau\leq0$; then Wilcoxon's signed rank statistic is $W=\sum
_{i=1}%
^{I}S_{i}q_{i}$, where the $q_{i}$ are a permutation of $1$, $2, \ldots
, I$.
Conditionally given $\mathcal{F},\mathcal{Z}$, if $H_{\tau}$ is true in a
randomized paired experiment, then $Y_{i}-\tau=\varepsilon_{i} $ is
$\pm\vert r_{Ci1}-r_{Ci2}\vert$\vspace*{1pt} with equal probabilities
$\frac{1}{2}$, so $q_{i}$ is fixed and $S_{i}=1$ or $0$ with equal
probabilities $\frac{1}{2}$, and Wilcoxon's statistic has the
distribution of
the sum of $I$ independent random variables taking the values $i$ or 0 with
equal probabilities $\frac{1}{2}$. This null distribution is the basis for
testing $H_{\tau}$, and by inverting the test it yields confidence intervals
and Hodges--Lehmann point estimates for an additive treatment effect
$\tau$.
See \citet{Leh75} for discussion of these standard techniques and for
discussion of the good performance of Wilcoxon's statistic when applied in
randomized experiments. See \citet{Mar79} for a parallel development of
randomization inferences using Huber's m-estimates including the permutational
$t$-test.

\subsection{\texorpdfstring{Sensitivity analysis in observational studies.}{Sensitivity analysis in observational studies}}

\label{ssReviewSen}

In the absence of randomization, there is no basis for assuming that
$\Pr( Z_{ij}=1\vert\mathcal{F},\mathcal{Z}) =\frac{1}{2} $ and
therefore no basis beyond na\"{\i}vet\'{e} for assuming that the
inferences in Section \ref{ssReviewRand} are correct. A~sensitivity
analysis in an observational study asks how the conclusions in
Section~\ref{ssReviewRand} would change in response to departures from
$\Pr( Z_{ij}=1\vert\mathcal{F},\mathcal{Z}) =\frac{1}{2}$ of various
magnitudes. One model for sensitivity analysis [\citet{Ros02a},
Section 4] begins by assuming that in the population before matching
treatment assignments are independent with unknown probabilities
$\pi_{ij}=\Pr( Z_{ij}=1\vert\mathcal{F}) $, and two subjects with, say,
$ij$ and $ij^{\prime}$, with the same observed covariates, $\mathbf{x}%
_{i1}=\mathbf{x}_{i2}$, may differ in their odds of treatment by at
most a
factor of $\Gamma\geq1$,%
%
\begin{equation}\label{eqSenMod}%
\frac{1}{\Gamma}\leq\frac{\pi_{ij}( 1-\pi_{ij^{\prime}}) }%
{\pi_{ij^{\prime}}( 1-\pi_{ij}) }\leq\Gamma\qquad\mbox{ if }%
\mathbf{x}_{i1}=\mathbf{x}_{i2};
\end{equation}
then a distribution of treatment assignments for matched pairs is
obtained by
conditioning on the event $\mathcal{Z}$. This is easily seen to be
equivalent to assuming that
%
\begin{equation}\label{eqSenMod2}
\frac{1}{1+\Gamma}\leq\Pr( Z_{i1}=1\vert\mathcal{F}%
,\mathcal{Z}) \leq\frac{\Gamma}{1+\Gamma},\qquad Z_{i2}=1-Z_{i1},\qquad
i=1,\ldots,I,\hspace*{-22pt}
\end{equation}
with independent assignments in distinct pairs; see
\citet{Ros02a}, Section~4. To aid interpretation, the one
parameter $\Gamma$ may be unpacked into two parameters, one $\Lambda$
controlling the relationship between treatment assignment $Z_{ij}$ and
the unobserved covariate $u_{ij}$, the other $\Delta$ controlling the
relation between response $r_{Cij}$ and $u_{ij}$, yielding the same
one-dimensional analysis in terms of $\Gamma$ but for all $(
\Lambda,\Delta) $ that solve $\Gamma=( \Lambda\Delta+1) /(
\Lambda+\Delta) $ [\citet{RosSil09}]; for instance, $\Gamma=1.25$
corresponds with an unobserved covariate that simultaneously doubles
the odds of treatment, $Z_{i1}-Z_{i2}=1 $, and doubles the odds of a
positive response difference under control, $r_{Ci1}-r_{Ci2}>0$, as
$1.25=( 2\times2+1) /( 2+2) $. In this formulation, the parameter
$\Delta$ is defined using Wolfe's (\citeyear{Wol74}) semiparametric
family of asymmetric deformations of a symmetric distribution to place
a bound on the distribution of $r_{Ci1}-r_{Ci2}$; see
\citet{RosSil09} for specifics. Either (\ref{eqSenMod}) or
(\ref{eqSenMod2}) says that treatment assignment probabilities are
unknown but to a bounded degree determined by $\Gamma$. For each fixed
$\Gamma\geq1$, (\ref{eqSenMod2}) yields an interval of possible values
of an inference quantity, such as a $P$-value or point estimate or the
endpoint of a confidence interval, and a sensitivity analysis consists
in computing that interval for several values of $\Gamma$, thereby
indicating the magnitude of departure from randomization that would
need to be present to alter the conclusions of the analysis in Section
\ref{ssReviewRand}. For instance, for $\Gamma=1$ the interval of
one-sided $P$-values from Wilcoxon's test is a single point, namely,
the $P$-value from the randomization test in Section~\ref{ssReviewRand},
but as $\Gamma\rightarrow\infty$ the interval tends
to $[ 0,1] $---that is, association does not logically imply causation.
The practical question is: how large must $\Gamma$ be before the
interval of $P$-values is inconclusive, say, including values both
above and below a conventional level such as 0.05?

Various methods of sensitivity analysis in observational studies are
discussed by \citet{CORetal59}, \citet{CopEgu01},
\citet{DipGan04}, \citet{EglSchMac09},
\citet{FraRub99}, \citet{Gas92}, \citet{GilBosHud03},
\citet{HosHanHol10}, \citet{Imb03},
\citet{LinPsaKro98}, \citet{Mar97},
\citet{McCGusLev07},
\citet{RosRub}, \citet{Sma07}, \citet{WanKri06},
\citet{Yan84}, \citet{YuGas05}, among others.

The discussion has emphasized adjustments for observed covariates by
matching, as opposed to, say, covariance adjustment. In simulations,
\citet{Rub79} found that model-based adjustments without matching
are not robust to model misspecification, sometimes increasing rather
than reducing bias from measured covariates, but he found that
model-based adjustments of matched pair differences are robust to model
misspecification. The methods described in the current paper may be
applied to residuals of covariance adjustment of matched pair
differences using the device in \citet{Ros02b}, Section 5. Also,
the sensitivity model (\ref{eqSenMod}) is applicable to a wide variety
of situations, including binary outcomes and censored survival times
[\citet{Ros02a}, Section 4].

\subsection{\texorpdfstring{Design sensitivity in observational studies.}{Design sensitivity in observational studies}}

\label{ssReviewDS}

If an observational study were free of unmeasured bias, then we could not
determine this from the observable data, and the best we could hope to
say is
that the conclusions are insensitive to small and moderate biases. The power
of a sensitivity analysis is the probability that we will be able to
say this
[\citet{Ros04}]. The power of a randomization test anticipates the outcome
of such a test under an assumed model for data generation in a randomized
trial. In parallel, the power of a sensitivity analysis with a specific
$\Gamma$ anticipates the outcome of a~sensitivity analysis when
performed on
data from an assumed model for data generation. In the \textit{favorable
situation}, the data reflect a treatment effect and no bias from unmeasured
covariates, and it is in this situation that we hope to report insensitivity
to unmeasured bias. For instance, we might ask the following: if the
$I$ matched pair
differences were produced by an additive constant treatment effect $\tau
$ with
no bias and Normal errors, $Y_{i}=\tau+\varepsilon_{i}$ with
$\varepsilon_{i}%
\sim_{\mathrm{i.i.d.}}N( 0,\sigma^{2}) $, then, under this model, what is the
probability that the entire interval of $P$-values testing $H_{0}$ is below
0.05 when computed with, say, $\Gamma=2$? For Wilcoxon's test with $I=100$
and $\tau/\sigma=1/2$, the entire interval of $P$-values computed with
$\Gamma=2$ is less than 0.05 with probability 0.54, so there is a reasonable
chance that an effect of this magnitude will be judged insensitive to a
moderately large bias of $\Gamma=2 $. In contrast, the new adaptive test
proposed in the current paper has power of 0.68 in this same situation, a
substantial improvement.

As $I\rightarrow\infty$, there is a value $\widetilde{\Gamma}$ called
the design sensitivity [\citet{Ros04}] such that the power of a
sensitivity analysis tends to 1 if the\vspace*{1pt} analysis is performed with
$\Gamma <\widetilde{\Gamma}$ and tends to zero if the analysis is
performed with $\Gamma>\widetilde{\Gamma}$. That is, the power, viewed
as a function of $\Gamma$ is tending to a~step function with a single
step down from 1 to 0 at $\Gamma=\widetilde{\Gamma}$; see
\citet{Ros10N2}, Figure 14.3. In the limit, as the sample size
increases, data generated by a certain model without bias will be
insensitive to biases smaller than $\widetilde{\Gamma}$ and sensitive
to biases larger than $\widetilde{\Gamma}$. For instance, if
$Y_{i}=\tau+\varepsilon_{i}$ with
$\varepsilon_{i}\sim_{\mathrm{i.i.d.}}N( 0,\sigma ^{2}) $ and
$\tau/\sigma=1/2$, the design sensitivity for Wilcoxon's statistic is
$\widetilde{\Gamma}=3.17$, so for sufficiently large $I$ it is
virtually certain that Wilcoxon's statistic will report insensitivity
to a bias of $\Gamma$ if $\Gamma<3.17$ and virtually certain it will
report sensitivity to a bias of $\Gamma$ if $\Gamma>3.17$. In contrast,
in this same sampling situation, the new adaptive test proposed in the
current paper has design sensitivity $\widetilde{\Gamma}=4.97$, again a
substantial improvement. In particular, in this sampling situation as
$I\rightarrow \infty$, the power of a sensitivity analysis performed at
$\Gamma=4$ is tending to zero for Wilcoxon's test and to one for the
new adaptive test.

Design sensitivity has been described in terms of the power of tests,
but parallel issues arise in conducting a sensitivity analysis for a
confidence interval or a point estimate. In a randomized experiment, a
test such as Wilcoxon's test\vadjust{\goodbreak} may be inverted to yield a confidence
interval or a Hodges--Lehmann point estimate of an additive treatment
effect $\tau$, and a more powerful test yields a typically shorter
confidence interval and more accurate point estimate; see
\citet{HodLeh63} or \citet{Leh75}, Section 4. In parallel, in
an observational study, a~sensitivity analysis for a~confidence
interval or a point estimate is obtained by inverting a test, so the
95\% confidence interval for a given $\Gamma$ excludes $\tau_{0}$ if
the sensitivity analysis for the test rejects
$H_{0}\dvtx\tau=\tau_{0}$; see Rosenbaum (\citeyear{Ros93},
\citeyear{Ros02a}), Section 4.3. For a~given $\Gamma>1$, one obtains an
interval of possible point estimates and a set of possible confidence
intervals. As $I\rightarrow\infty$ with $\Gamma>1$ fixed, both the
interval of possible point estimates and the union of possible
confidence intervals converges to a real interval $[ \tau_{L},\tau_{H}]
$ of the treatment effects $\tau$ that are compatible with a bias of
$\Gamma$, and an increase in design sensitivity will shorten that
interval; see Rosenbaum [(\citeyear{Ros05}), Proposition 1] for one
such result. As in experiments, even if one is interested in a
confidence interval or point estimate, not a hypothesis test, one
should obtain that interval or estimate by inverting a more powerful
test.

\section{\texorpdfstring{Why is adaptation important\textup{?}}{Why is adaptation
important}}

\label{ssWhyImportant}

Traditionally, adaptive methods have selected the best of several
statistical procedures using the data at hand and they have focused on
improving efficiency in randomized experiments in the absence of bias;
see, for instance, \citet{Hog74}, \citet{PolHet76} and
\citet{Jon79}. As discussed in Rosenbaum (\citeyear{Ros10N1},
\citeyear{Ros}), Pitman efficiency and design sensitivity both affect
the power of a sensitivity analysis in an observational study, but they
can work at cross-purposes. Pitman efficiency aims at power to detect
small effects in randomized experiments where bias is not an issue. In
an observational study, Pitman efficiency predicts the outcome of a
sensitivity analysis for $\Gamma=1$ in the favorable situation; that
is, it predicts the outcome of a randomization test applied in an
observational study when bias is eliminated by adjustments such as
matching. Small effects, however, are invariably sensitive to small
unobserved biases, which are absent in an idealized randomized
experiment but can never be excluded from consideration in an
observational study. Procedures with superior design sensitivity in
observational studies look for stable evidence of moderately large
effects, in effect ignoring pairs $i$ with small $\vert Y_{i}\vert$.

There are procedures with good Pitman efficiency and better design
sensitivity than Wilcoxon's statistic, and other statistics with poor
Pitman efficiency and vastly better design sensitivity than Wilcoxon's
statistic. For instance, in testing $H_{0}$, \citet{Bro81}
proposed a statistic which ignores the $\frac{1}{3}$ of pairs with the
smallest $\vert Y_{i}\vert$ or $q_{i}$, gives weight 1 to the signs of
the $\frac{1}{3}$ of pairs with the middle values of $\vert Y_{i}\vert$
or $q_{i}$, and gives weight 2 to the remaining $\frac{1}{3}$ of pairs
with the largest values of $\vert Y_{i}\vert$ or $q_{i}$.
\citet{Bro81} shows his statistic is highly robust and almost as
efficient as Wilcoxon's statistic in a randomized experiment, whereas
in \citet{Ros10N1} it is seen that Brown's statistic has higher
design sensitivity in a range of sampling situations; in combination,
these two facts produce improved power in a sensitivity analysis.
\citet{Noe73} proposed a simpler class of statistics that simply
counts the number of positive $Y_{i}$ among pairs with large~$\vert
Y_{i}\vert$ or $q_{i}$. \citet{MarHet82} studied the Pitman
efficiency of many statistics similar to those of Brown and Noether by
varying the number of pairs that are given various weights; see also
the group rank statistics of \citet{Gas66} and Groeneveld
(\citeyear{Groeneveld72}). In the version used in the current paper---but not
in Noether's paper---Noether's statistic counts the number of positive $Y_{i}$
among the $\frac{1}{3}$ of pairs with largest $\vert Y_{i}\vert$ or
$q_{i}$. Having mentioned once that Noether did not promote this
specific version of his statistic, I will not mention this again, and
will refer to the statistic as Noether's statistic. Brown's statistic
has Pitman efficiency of 0.95 relative to the Wilcoxon statistic for an
additive effect with Normal errors, but Noether's statistic has Pitman
efficiency of only 0.78, so one would not use this version of Noether's
statistic for Normal data from a randomized experiment. Table
\ref{tabeff} gives Pitman efficiencies in a paired randomized
experiment. In contrast, in a sensitivity analysis in an observational
study, if $Y_{i}=\tau +\varepsilon_{i}$ with
$\varepsilon_{i}\sim_{\mathrm{i.i.d.}}N( 0,\sigma^{2}) $ and $\tau
/\sigma=1/2$, the design sensitivity for Wilcoxon's statistic is
$\widetilde{\Gamma}=3.17$, for Brown's statistic is $\widetilde{\Gamma
}=3.60 $ and for Noether's statistic is $\widetilde{\Gamma}=4.97$, so
for sufficiently large $I$ Noether's statistic will be the best
performer in a
sensitivity analysis.%

%

\begin{table}
\tablewidth=270pt
\caption{Pitman asymptotic relative efficiency versus the Wilcoxon
statistic for a shift alternative in a paired randomized experiment
with errors from a Normal distribution, a logistic distribution or a
$t$-distribution with 3~degrees of freedom. In the version used here,
Noether's statistic is the~number of positive differences among the $1/3$
of pairs with the largest~absolute differences} \label{tabeff}
\begin{tabular*}{\tablewidth}{@{\extracolsep{\fill}}l ccc@{}}
\hline
& \textbf{Normal} & \textbf{Logistic} & $\bolds{t}$ \textbf{3 df} \\
\hline
Sign & 0.67 & 0.75 & 0.85 \\
Noether & 0.78 & 0.69 & 0.59 \\
Brown & 0.95 & 0.94 & 0.93 \\
Wilcoxon & 1.00 & 1.00 & 1.00 \\
\hline
\end{tabular*}
\end{table}
%

The adaptive test uses both Brown's statistic and Noether's statistic.
Adjusting the critical values to control the level of the test, the adaptive
test rejects if either Brown's statistic or Noether's statistic supports
rejection. In every sampling situation, the adaptive test has the
larger of
the two design sensitivities for Brown's and Noether's statistics. The
important issue, however, is the power of a sensitivity analysis for finite
$I$. Because asymptotic claims for some adaptive procedures are not readily
seen in samples of plausible size, the current paper uses the exact null
distribution of the adaptive test and emphasizes finite sample power
determined by simulation. The pairing of Brown's statistic and Noether's
statistic is a pairing of two strong candidates for which the required exact
calculations are feasible.

\section{\texorpdfstring{An adaptive test.}{An adaptive test}}

\label{secAdaptive}

\subsection{\texorpdfstring{The exact null distribution in a sensitivity analysis.}{The exact null distribution in a sensitivity
analysis}}

\label{ssExactNulol}

Let $0\leq\lambda_{1}<\lambda_{2}\leq1$. Let $I_{1}$ be the number of pairs
with absolute ranks $q_{i}\geq( 1-\lambda_{1}) I$ and let $B_{1}$
be the number of positive $Y_{i}$ among these $I_{1}$ pairs. Also, let
$I_{2}$ be the number of ranks with $( 1-\lambda_{1}) I>q_{i}%
\geq( 1-\lambda_{2}) I$ and let $B_{2}$ be the number of positive
$Y_{i}$ among these $I_{2}$ pairs. \citet{Noe73} proposed $B_{1}$
as a test statistic, and \citet{Bro81} and \citet{MarHet82}
proposed $T=2B_{1}+B_{2}$ as a test statistic; see also
\citet{Gas66}.

Let $\overline{\overline{B}}_{1}$ and $\overline{\overline{B}}_{2}$ be
independent binomials with sample sizes $I_{1}$ and $I_{2}$ and probabilities
of success $\kappa=\Gamma/( 1+\Gamma) $, and let $\overline
{B}_{1}$ and $\overline{B}_{2}$ be independent binomials with sample sizes
$I_{1}$ and $I_{2}$ and probabilities of success \mbox{$\kappa=1/(
1+\Gamma) $}. Also, let $\overline{\overline{T}}=2\overline
{\overline{B}}_{1}+\overline{\overline{B}}_{2}$ and $\overline{T}%
=2\overline{B}_{1}+\overline{B}_{2}$. A function $g( \cdot
,\cdot) $ is monotone increasing if $g( b_{1},b_{2}) \leq
g( b_{1}^{{\prime}},b_{2}^{{\prime}}) $ whenever $b_{1}\leq
b_{1}^{{\prime}}$ and $b_{2}\leq b_{2}^{{\prime}}$. Under the sensitivity
model~(\ref{eqSenMod2}), if $H_{0}$ is true, then it is not difficult
to show
[\citet{Ros02a}, Section 4] that for every monotone increasing
function~$g(\cdot,\cdot), $%
%
\begin{eqnarray}\label{eqSenBound}%
\Pr\{ g( \overline{B}_{1},\overline{B}_{2}) \geq
k \} &\leq&\Pr\{ g( B_{1},B_{2}) \geq
k \vert \mathcal{F}, \mathcal{Z}\} \nonumber\\[-8pt]\\[-8pt]
&\leq&\Pr\{ g(
\overline{\overline{B}}_{1},\overline{\overline{B}}_{2}) \geq
k \} \qquad\mbox{for every }k,\nonumber
\end{eqnarray}
and the bounds in (\ref{eqSenBound}) are sharp in the sense of being attained
for some $\Pr( Z_{i1}=1 \vert \mathcal{F}%
, \mathcal{Z}) $ that satisfy (\ref{eqSenMod2}), so the bounds
(\ref{eqSenBound}) cannot be improved without additional information that
further restricts $\Pr( Z_{i1}=1 \vert \mathcal{F}%
, \mathcal{Z}) $. If \mbox{$\Gamma=1$}, then there is equality throughout
(\ref{eqSenBound}) and then (\ref{eqSenBound}) is the randomization
distribution of $g( B_{1},B_{2}) $ under~$H_{0}$.

Let $g( B_{1},B_{2}) =1$ if $B_{1}\geq k_{B,\Gamma}$ or
$2B_{1}+B_{2}\geq k_{T,\Gamma}$ and $g( B_{1},B_{2}) =0$
otherwise, for suitable constants $k_{B,\Gamma}$ and $k_{T,\Gamma}$; then
$g( \cdot,\cdot) $ is monotone increasing. For a given
$\Gamma\geq1$, the adaptive test rejects $H_{0}$ at level $\alpha$ for all
$\pi_{ij}$ satisfying (\ref{eqSenMod}) if $B_{1}\geq k_{B,\Gamma}$ or
$2B_{1}+B_{2}\geq k_{T,\Gamma}$. The constants $k_{B,\Gamma}$ and
$k_{T,\Gamma}$ are determined to satisfy the following conditions:%
%
\begin{eqnarray}
\label{eqCrit1}%
\Pr( \overline{\overline{B}}_{1}\geq k_{B,\Gamma} \mbox{ or }%
\overline{\overline{T}}\geq k_{T,\Gamma}) &\leq&\alpha,
\\
\label{eqCrit2}%
\Pr( \overline{\overline{B}}_{1}\geq k_{B,\Gamma}-1 \mbox{ or }%
\overline{\overline{T}}\geq k_{T,\Gamma}) &>&
\alpha\quad\mbox{and}\nonumber\\[-8pt]\\[-8pt]
\Pr( \overline{\overline{B}}_{1}\geq k_{B,\Gamma} \mbox{ or }%
\overline{\overline{T}}\geq k_{T,\Gamma}-1) &>&\alpha\nonumber
\end{eqnarray}
and
%
\begin{equation}\label{eqCrit3}\quad
\vert{\Pr}( \overline{\overline{B}}_{1}\geq k_{B,\Gamma})
-\Pr( \overline{\overline{T}}\geq k_{T,\Gamma}) \vert
\quad\mbox{is minimized subject to (\ref{eqCrit1}) and (\ref{eqCrit2}%
).}\vadjust{\goodbreak}
\end{equation}
The joint distribution of $( \overline{\overline{B}}_{1},\overline
{\overline{B}}_{2}) $ is that of two independent binomials and in
\textsf{R} is given by \textsf{outer}(\textsf{dbinom}(0: $I_{1}$,
$I_{1}$, $\kappa$), \textsf{dbinom}(0: $I_{2}$, $I_{2}$,
$\kappa$), ``*''); then finding $k_{B,\Gamma}$ and $k_{T,\Gamma}$ to
satisfy (\ref{eqCrit1})--(\ref{eqCrit3}) is simply arithmetic.

Although I have never seen this, in principle, there could be two
values, $( k_{B,\Gamma},k_{T,\Gamma}) $ and $( k_{B,\Gamma
}^{{\prime}},k_{T,\Gamma}^{{\prime}}) =( k_{B,\Gamma
}-1, k_{T,\Gamma}+1) $ that both satisfy (\ref{eqCrit1}%
)--(\ref{eqCrit3}). To avoid this ambiguity in the definition of the adaptive
procedure, simply use $( k_{B,\Gamma},k_{T,\Gamma}) $ in this
extremely unlikely case, thereby preferring to reduce the critical
value for
Brown's statistic $T$.

\subsection{\texorpdfstring{Numerical example of the null distribution.}{Numerical example of the null
distribution}}

\label{ssNumericalNull}

To illustrate the computations in (\ref{eqCrit1})--(\ref{eqCrit3}), take
$I=250$ untied pairs, $\Gamma=4$, $\alpha=0.05$, and $\lambda_{1}=1/3$,
$\lambda_{2}=2/3$; then, $I_{1}=84$, $I_{2}=83$, $\kappa=4/5$. This yields
$k_{B,\Gamma}=74$ and $k_{T,\Gamma}=216$ with
%
\begin{eqnarray}
\label{eqCritEg1}%
&\Pr( \overline{\overline{B}}_{1}\geq74 \mbox{ or } \overline
{\overline{T}}\geq216) = 0.0488\leq\alpha=0.05,&
\\
\label{eqCritEg2}%
&\Pr( \overline{\overline{B}}_{1}\geq74)=0.0370,\qquad
\Pr( \overline{\overline{T}}\geq216) =0.0320,&
\\
\label{eqCritEg3}%
&\vert{\Pr}( \overline{\overline{B}}_{1}\geq74) -\Pr(
\overline{\overline{T}}\geq216) \vert = 0.0050.&
\end{eqnarray}
In light of this, for $\Gamma=4$, the upper bound on the one-sided $P$-value
testing no effect would be less than $\alpha=0.05$ if either $B_{1}\geq
74$ or
$T=2B_{1}+B_{2}\geq216$.

Several aspects of the illustration (\ref{eqCritEg1})--(\ref{eqCritEg3})
deserve comment. First, if one were to test using $B_{1}$ alone, ignoring
$B_{2}$, then at $\Gamma=4$ the upper bound on the one-sided $P$-value testing
no effect would be less than $\alpha=0.05$ if $B_{1}\geq74$, because
$\Pr( \overline{\overline{B}}_{1}\geq74) =0.0370$ and $\Pr(
\overline{\overline{B}}_{1}\geq73) =0.0691$; that is, in this
particular case, owing to the discreteness of the binomial
distribution, the
adaptive test will reject in every instance in which the test based on~$B_{1}$
alone rejects and the adaptive test will reject in some other cases as well.
Conversely, if one were to test using $T$ alone, then at $\Gamma=4$ the
upper bound on the one-sided $P$-value testing no effect would be less than
$\alpha=0.05$ if $T\geq215$ rather than $k_{T,\Gamma}=216$ in (\ref
{eqCritEg1}%
) because $\Pr( \overline{\overline{T}}\geq215) =0.04288$ and
$\Pr( \overline{\overline{T}}\geq214) =0.05642$. So, in this
one numerical example, the adaptive test rejects in every instance in which
the test based on $B_{1}$ rejects and also in every instance in which $T$
rejects except $B_{1}<74$ and $T=215$. Use of the Bonferroni inequality to
approximate $\Pr( \overline{\overline{B}}_{1}\geq74 \mbox{ or }
\overline{\overline{T}}\geq216) $ would err substantially, with
$\Pr( \overline{\overline{B}}_{1}\geq74 \mbox{ or } \overline
{\overline{T}}\geq216) =$ $0.0488\leq\Pr( \overline{\overline{B}%
}_{1}\geq74) +\Pr( \overline{\overline{T}}\geq216) $
$=0.0370+0.0320=0.0690$.

\subsection{\texorpdfstring{Design sensitivity of the adaptive test.}{Design sensitivity of the adaptive
test}}

\label{ssAdaptiveDS}

As discussed in Section \ref{ssReviewDS}, in an observational study, the
\textit{favorable situation} means there is a treatment effect and no bias
from an unmeasured covariate. In an observational study, we cannot know from
the data whether we are in the favorable situation, so the best we can
hope to
say is that the study's conclusions are insensitive to small and moderate
biases. The power of an $\alpha$-level sensitivity analysis, $0<\alpha<1$,
performed with a specific \mbox{$\Gamma\geq1$}, is the probability that the entire
interval of possible $P$-values from the sensitivity analysis is less
than or
equal to $\alpha$. For the adaptive test, the power of the sensitivity
analysis for fixed $\Gamma$ is the probability that $B_{1}\geq
k_{B,\Gamma
} \mbox{ or } T\geq k_{T,\Gamma}$ when $B_{1}$ and $T$ are computed from
data that are, in fact, measuring a treatment effect without bias. In
principal, one could compute the power conditional upon $\mathcal{F}$, but
this would mean that the power would be a function of $\mathcal{F}$,
so, in
practice, one computes the unconditional power averaging over a simple model
for the generation of $\mathcal{F}$. As noted in Section \ref
{ssReviewDS}, the
design sensitivity is a number $\widetilde{\Gamma}$ such that the power
of an
$\alpha$-level sensitivity\vspace*{1pt} analysis tends to 1 as $I\rightarrow\infty$
if the
sensitivity analysis is performed with $\Gamma<\widetilde{\Gamma}$ and the
power tends to zero if the analysis is performed with $\Gamma
>\widetilde{\Gamma}$.

In the current paper, the favorable situation refers to treated-minus-control
differences $Y_{i}$ that are drawn independently from a continuous cumulative
distribution $F( \cdot) $ that is strictly increasing, $F(
y) <F( y^{\prime}) $ if $y<y^{\prime}$. One of many such
favorable situations is $Y_{i}=\tau+\varepsilon_{i}$ where the
$\varepsilon_{i}$ are
independent and identically distributed observations from a continuous,
strictly increasing distribution with a density symmetric about zero.

For $y\geq0$, let $H(y)=F( y) -F( -y) $; then
$H( y) =\Pr( \vert Y_{i}\vert\leq y) $,
and for $\lambda\in[ 0,1) $, the inverse function is well defined
with $H^{-1}( \lambda) =y$ if $\lambda=\Pr( \vert
Y_{i}\vert\leq y) $. Also define $\zeta( \lambda)
$ to be the probability that a $Y_{i}$ is both positive, $Y_{i}>0$, and
in the
largest $\lambda$ of the $\vert Y_{i}\vert$, that is, define
\[
\zeta( \lambda) =1-F\{ H^{-1}( 1-\lambda)
\} =\Pr[ ( Y_{i}>0) \wedge\{ \vert
Y_{i}\vert>H^{-1}( 1-\lambda) \} ].
\]
Let $\widetilde{\Gamma}_{\mathrm{no}}$, $\widetilde{\Gamma}_{\mathrm{bmh}}$
and $\widetilde{\Gamma}_{\mathrm{ad}}$ be the design sensitivities for,
respectively, Noether's statistic $B_{1}$, the Brown--Markowski--Hettmansperger
statistic $T$ and the adaptive procedure with critical values (\ref
{eqCrit1}%
)--(\ref{eqCrit3}). That is, $B_{1}$ counts the positive $Y_{i}$'s among the
largest $\lambda_{1}$ of the $\vert Y_{i}\vert$, and
$T=2B_{1}+B_{2}$ doubles $B_{1}$ and adds the count of the positive
$Y_{i}$'s among the next $\lambda_{2}-\lambda_{1}$ of the $\vert
Y_{i}\vert$.
%
\begin{proposition}
\label{PropDS}
$\!\!\!$If $Y_{i}, i\,{=}\,1,\ldots,I$ are independent observations from~$F( \cdot) $,
%
\begin{eqnarray}
\label{eqDSno}%
\widetilde{\Gamma}_{\mathrm{no}}&=&\frac{\zeta( \lambda_{1})
}{\lambda_{1}-\zeta( \lambda_{1}) },
\\
\label{eqDSbrown}%
\widetilde{\Gamma}_{\mathrm{bmh}}&=&\frac{\zeta( \lambda_{1})
+\zeta( \lambda_{2}) }{\{ \lambda_{1}-\zeta(
\lambda_{1}) \} +\{ \lambda_{2}-\zeta( \lambda
_{2}) \} },
\\
\label{eqDSadapt}%
\widetilde{\Gamma}_{\mathrm{ad}}&=&\max( \widetilde{\Gamma}_{\mathrm{no}%
}, \widetilde{\Gamma}_{\mathrm{bmh}}).
\end{eqnarray}
\end{proposition}
\begin{pf}
The proof uses Proposition 2 of \citet{Ros10N1} which concerns the
design sensitivity of a signed rank statistic with general scores; in
particular,\vadjust{\goodbreak} $B_{1}$ and $T$ are two such signed rank statistics.
Equations~(\ref{eqDSno}) and~(\ref{eqDSbrown}) are obtained by
simplifying expression (8) in Proposition 2 of \citet{Ros10N1},
which is a formula for the design sensitivity with general scores. As
shown in the proof of that proposition, in a sensitivity analysis
performed at a specific value of $\Gamma$, the upper bound on the
$P$-value for $B_{1}$ converges in probability to zero as
$I\rightarrow\infty$ if $\Gamma<\widetilde{\Gamma}_{\mathrm{no}} $ and
it converges to 1 if $\Gamma>\widetilde{\Gamma}_{\mathrm{no}}$, and, in
parallel, the upper bound on the $P$-value for $T$ converges in
probability to zero as $I\rightarrow\infty$ if
$\Gamma<\widetilde{\Gamma}_{\mathrm{bmh}} $ and it converges to 1 if
$\Gamma>\widetilde{\Gamma}_{\mathrm{bmh}}$. As a consequence, the
smaller of these two $P$-values for $B_{1}$ and $T$ tends to zero as
$I\rightarrow\infty$ if $\Gamma<\max( \widetilde{\Gamma
}_{\mathrm{no}}, \widetilde{\Gamma}_{\mathrm{bmh}}) $ and it tends to 1
if $\Gamma>\max( \widetilde{\Gamma}_{\mathrm{no}}, \widetilde{\Gamma
}_{\mathrm{bmh}}) $, proving (\ref{eqDSadapt}).
\end{pf}

Table \ref{tabds} calculates the design sensitivity of the sign statistic, the Wilcoxon
signed rank statistic, Noether's statistic with $\lambda_{1}=1/3$, Brown's
statistic with $\lambda_{1}=1/3$, $\lambda_{2}=2/3$, and the adaptive
procedure with critical values~(\ref{eqCrit1})--(\ref{eqCrit3}). In
%
%
\begin{table}
\tablewidth=300pt
\caption{Design sensitivity $\widetilde{\Gamma }$ in the favorable
situation with an additive treatment effect, $\tau$~and errors
$\varepsilon_{i}$ with variance $\sigma^{2}$ that are Normal, logistic
or $t$-distributed with~3~degrees of freedom}
\label{tabds}
\begin{tabular*}{\tablewidth}{@{\extracolsep{\fill}}l d{2.2}d{2.2}d{2.2}@{}}
\hline
\textbf{Statistic} & \multicolumn{1}{c}{\textbf{Normal}}
& \multicolumn{1}{c}{\textbf{Logistic}}
& \multicolumn{1}{c@{}}{$\bolds{t}$ \textbf{3 df}} \\
\hline
& \multicolumn{3}{c@{}}{${\tau/\sigma=1/4}$} \\[4pt]
Sign & 1.49 & 1.57 & 1.88 \\
Wilcoxon & 1.76 & 1.83 & 2.21 \\
Brown & 1.86 & 1.93 & 2.34 \\
Noether & 2.12 & 2.14 & 2.48 \\
Adaptive & 2.12 & 2.14 & 2.48 \\
[4pt]
& \multicolumn{3}{c@{}}{$\tau/\sigma=1/2$} \\
[4pt]
Sign & 2.24 & 2.48 & 3.44 \\
Wilcoxon & 3.17 & 3.40 & 4.74 \\
Brown & 3.60 & 3.83 & 5.39 \\
Noether & 4.97 & 4.72 & 5.77 \\
Adaptive & 4.97 & 4.72 & 5.77 \\
[4pt]
& \multicolumn{3}{c@{}}{$\tau/\sigma=3/4$} \\
[4pt]
Sign & 3.41 & 3.90 & 6.02 \\
Wilcoxon & 5.92 & 6.42 & 9.70 \\
Brown & 7.55 & 7.91 & 11.69 \\
Noether & 13.48 & 10.86 & 12.08 \\
Adaptive & 13.48 & 10.86 & 12.08 \\
\hline
\end{tabular*}
\end{table}
Table \ref{tabds},
$Y_{i}=\tau+\varepsilon_{i}$ where $\operatorname{var}( \varepsilon_{i})
=\sigma^{2}$, the effect size is specified in units of the standard deviation,
$\tau/\sigma$, and $\varepsilon_{i}$ has a standard Normal
distribution, a
logistic distribution or a central $t$-distribution with 3 degrees of freedom.
For example, if one takes $\sigma=1$, then for the Normal $\tau/\sigma=1/2$
if $\tau=1/2$, for the logistic $\tau/\sigma=1/2$ if $\tau=( 1/2)
( \pi/\sqrt{3}) \doteq0.907$, and for the $t$-distribution with 3
degrees of freedom, $\tau/\sigma=1/2$ if $\tau=( 1/2) \sqrt
{3}\doteq0.866$. Although $\widetilde{\Gamma}_{\mathrm{no}}%
>\widetilde{\Gamma}_{\mathrm{bmh}}$ throughout Table \ref{tabds}, there are many
situations with $\widetilde{\Gamma}_{\mathrm{no}}<\widetilde{\Gamma
}_{\mathrm{bmh}}$; for instance, with $\lambda_{1}=1/3$ this can occur
in a $t$-distribution with 2 or 1 degrees of freedom, where the $t$
with 1 degree of freedom is the Cauchy distribution, and it occurs in
the $t$-distribution with 3 degrees of freedom in Table
\ref{tabcompare2} of Section \ref{ssDiscVariations} with
$\lambda_{1}<1/3$.%

As an illustration of the properties of design sensitivity, consider
the case
of $\tau/\sigma=1/2$ in Table \ref{tabds} for the $t$-distribution with 3 degrees of
freedom. The design sensitivity for Wilcoxon's statistic in this case is
$\widetilde{\Gamma}=4.74$, whereas for Noether's statistic it is
$\widetilde{\Gamma}_{\mathrm{no}}=5.77$. In sufficiently large samples from
this distribution, Wilcoxon's statistic should be sensitive to a bias of
magnitude $\Gamma=5$ but Noether's statistic should not. Drawing a single
sample of $I=10\mbox{,}000$ pairs from this distribution and performing a sensitivity
analysis with $\Gamma=5$ yields an upper bound on the $P$-value for Wilcoxon's
statistic of 0.9985 and for Noether's statistic of 0.0071, so a
deviation from
random assignment of magnitude $\Gamma=5$ could readily explain the observed
value of Wilcoxon's statistic, but not the observed value of Noether's
statistic. At the $\alpha=0.011$ level with $\Gamma=5$, the adaptive test
rejects $H_{0}$ because Noether's statistic has passed its critical
point in
(\ref{eqCrit1}) although Brown's statistic has not. Because~$I$ was very
large in this illustration, test performance was predicted by the design
sensitivity, but in smaller sample sizes, both design sensitivity and
efficiency affect test performance.

\subsection{\texorpdfstring{Simulation: Power of a sensitivity analysis in the favorable
situation.}{Simulation: Power of a sensitivity analysis in the favorable
situation}}
\label{ssSimulation}

The power of a sensitivity analysis is examined for finite $I$ by simulation
in Tables \ref{tabsimhalf} and \ref{tabsimquarter}. The tables describe the favorable situation:
there is a~treatment effect and no bias from unobserved covariates, but of course the
investigator does not know this in an observational study, and so
performs a~sensitivity analysis. The power of a 0.05-level sensitivity analysis is the
probability that the upper bound on the one-sided $P$-value is less
than 0.05.
The power is determined for Wilcoxon's signed rank test, Brown's test,
Noether's test and the adaptive test that uses both Brown's and Noether's
tests.

\begin{table}
\caption{Simulated power with $I$ pairs of a 0.05 level sensitivity
analysis performed with sensitivity parameter $\Gamma $. In each
situation, there is no bias and there is an additive constant treatment
effect $\tau $ whose magnitude is half the standard deviation of the
pair differences $Y_{i}$, so $\tau/\sigma = 1/2$. Each situation is
replicated 10,000 times. In each comparison, the two highest powers are
in \textup{bold}}
\label{tabsimhalf}
\begin{tabular*}{\tablewidth}{@{\extracolsep{\fill}}l ccc ccc ccc@{}}
\hline
\multicolumn{1}{@{\hspace*{-5pt}}l}{\hspace*{30pt}\textbf{Pairs}\textbf{:}\hspace*{-20pt}}
& \multicolumn{3}{c}{$\bolds{I=100}$} & \multicolumn{3}{c}{$\bolds{I=250}$}
& \multicolumn{3}{c@{}}{$\bolds{I=500}$}\\[-4pt]
& \multicolumn{3}{c}{\hrulefill} & \multicolumn{3}{c}{\hrulefill}
& \multicolumn{3}{c@{}}{\hrulefill}\\
\multicolumn{1}{@{\hspace*{-5pt}}l}{\hspace*{44pt}$\bolds{\Gamma}$\textbf{:}\hspace*{-20pt}}
& \multicolumn{1}{c}{\textbf{1}} & \multicolumn{1}{c}{\textbf{2}}
& \multicolumn{1}{c}{\textbf{3}} & \multicolumn{1}{c}{\textbf{2}}
& \multicolumn{1}{c}{\textbf{3}} & \multicolumn{1}{c}{\textbf{4}}
& \multicolumn{1}{c}{\textbf{3}} & \multicolumn{1}{c}{\textbf{4}}
& \multicolumn{1}{c@{}}{\textbf{5}} \\
\hline
& \multicolumn{9}{c}{Normal errors, $\tau/\sigma= 1/2$}\\
[4pt]
Wilcoxon & \textbf{1.00}
& 0.53 & 0.05 & 0.89 & 0.07 & 0.00 & 0.10 & 0.00 & 0.00 \\
Brown & \textbf{1.00}
& 0.61 & 0.10 & 0.93 & 0.19 & 0.01 & 0.35 & 0.01 & 0.00 \\
Noether & 0.99 & \textbf{0.64} & \textbf{0.29} & \textbf{0.96}
& \textbf{0.55} & \textbf{0.15} & \textbf{0.81} & \textbf{0.24}
& \textbf{0.03} \\
Adaptive & 1.00 & \textbf{0.68} & \textbf{0.17} & \textbf{0.97}
& \textbf{0.45} & \textbf{0.15} & \textbf{0.76} & \textbf{0.24}
& \textbf{0.03} \\
[4pt]
& \multicolumn{9}{c@{}}{Logistic errors, $\tau/\sigma= 1/2$}\\
[4pt]
Wilcoxon & \textbf{1.00}
& 0.65 & 0.10 & 0.95 & 0.16 & 0.00 & 0.25 & 0.00 & 0.00 \\
Brown & \textbf{1.00} & \textbf{0.70} & 0.15 & \textbf{0.97}
& 0.28 & 0.02 & 0.53 & 0.03 & 0.00 \\
Noether & 0.99 & 0.61 & \textbf{0.26} & 0.95 & \textbf{0.48} & \textbf
{0.11} & \textbf{0.75} & \textbf{0.17} & \textbf{0.02} \\
Adaptive & 1.00 & \textbf{0.70} & \textbf{0.18} & \textbf{0.97}
& \textbf{0.41} & \textbf{0.11} & \textbf{0.70} & \textbf{0.17}
& \textbf{0.02} \\
[4pt]
& \multicolumn{9}{c@{}}{$t$ errors with 3 d.f., $\tau/\sigma= 1/2$}\\
[4pt]
Wilcoxon & \textbf{1.00}
& \textbf{0.94} & 0.45 & \textbf{1.00}
& 0.81 & 0.21 & 0.98 & 0.33 & 0.02 \\
Brown & \textbf{1.00} & \textbf{0.94} & \textbf{0.48} & \textbf{1.00}
& \textbf{0.86} & \textbf{0.37} & \textbf{0.99} & \textbf{0.62} & 0.10
\\
Noether & 1.00 & 0.77 & 0.42 & 0.99 & 0.75 & 0.29 & 0.95 & 0.50 &
\textbf
{0.13} \\
Adaptive & 1.00 & 0.92 & \textbf{0.49} & 1.00 & \textbf{0.87} & \textbf
{0.37} & \textbf{0.99} & \textbf{0.58} & \textbf{0.14} \\
\hline
\end{tabular*}
\end{table}
\begin{table}
\caption{Simulated power with $I$ pairs of a 0.05 level sensitivity
analysis performed with sensitivity parameter $\Gamma $. In each
situation, there is no bias and there is an additive constant treatment
effect $\tau $ whose magnitude is either $1/4$ or $3/4$ of the standard
deviation $\sigma$ of the pair differences $Y_{i}$, so $\tau /\sigma=
1/4$ or $\tau/\sigma= 3/4$. Each situation is replicated 10,000 times.
In each comparison, the two highest powers are in \textup{bold}}
\label{tabsimquarter}
\begin{tabular*}{\tablewidth}{@{\extracolsep{\fill}}l cc cc cc cc@{}}
\hline
& \multicolumn{4}{c}{$\bolds{\tau/\sigma= 1/4}$} &
\multicolumn{4}{c@{}}{$\bolds{\tau/\sigma= 3/4}$} \\[-4pt]
& \multicolumn{4}{c}{\hrulefill} & \multicolumn{4}{c@{}}{\hrulefill} \\
\multicolumn{1}{@{}l}{\hspace*{30pt}\textbf{Pairs}\textbf{:}\hspace*{-20pt}}
& \multicolumn{2}{c}{$\bolds{I=100}$} & \multicolumn{2}{c}{$\bolds{I=500}$}
& \multicolumn{2}{c}{$\bolds{I=100}$} & \multicolumn{2}{c@{}}{$\bolds{I=500}$}\\[-4pt]
& \multicolumn{2}{c}{\hrulefill} & \multicolumn{2}{c}{\hrulefill} &
\multicolumn{2}{c}{\hrulefill} & \multicolumn{2}{c@{}}{\hrulefill}\\
\multicolumn{1}{@{}l}{\hspace*{44pt}$\bolds{\Gamma}$\textbf{:}\hspace*{-20pt}}
& \multicolumn{1}{c}{\textbf{1}} & \multicolumn{1}{c}{\textbf{1.5}}
& \multicolumn{1}{c}{\textbf{1.5}} & \multicolumn{1}{c}{\textbf{1.75}}
& \multicolumn{1}{c}{\textbf{2.5}} & \multicolumn{1}{c}{\textbf{3.5}}
& \multicolumn{1}{c}{\textbf{5}} & \multicolumn{1}{c@{}}{\textbf{6}}\\
\hline
& \multicolumn{8}{c}{Normal errors} \\
[4pt]
Wilcoxon & \textbf{0.78} & 0.16 & 0.44 & 0.05 & 0.92 & 0.49 & 0.28 & 0.02 \\
Brown & \textbf{0.75} & 0.18 & 0.53 & 0.11 & 0.94 & 0.67 & 0.78 & 0.33
\\
Noether & 0.60 & \textbf{0.20} & \textbf{0.65} & \textbf{0.33} & \textbf{0.97}
& \textbf{0.78} & \textbf{0.99} & \textbf{0.92} \\
Adaptive & 0.72 & \textbf{0.22} & \textbf{0.67} & \textbf{0.28} &
\textbf
{0.96} & \textbf{0.80} & \textbf{0.99} & \textbf{0.87} \\
[4pt]
& \multicolumn{8}{c@{}}{Logistic errors} \\
[4pt]
Wilcoxon & \textbf
{0.83} & 0.20 & 0.59 & 0.11 & 0.95 & 0.60 & 0.51 & 0.08 \\
Brown & \textbf{0.79} & \textbf{0.21} & \textbf{0.66} & 0.18 & \textbf{0.95}
& \textbf{0.71} & 0.85 & 0.42 \\
Noether & 0.60 & 0.19 & 0.65 & \textbf{0.33} & 0.93 & 0.67 & \textbf{0.93}
& \textbf{0.75} \\
Adaptive & 0.76 & \textbf{0.23} & \textbf{0.70} & \textbf{0.29} &
\textbf
{0.96} & \textbf{0.73} & \textbf{0.94} & \textbf{0.68} \\
[4pt]
& \multicolumn{8}{c@{}}{$t$ errors, 3 d.f.} \\
[4pt]
Wilcoxon &
\textbf
{0.96} & \textbf{0.47} & \textbf{0.97} & 0.67 & \textbf{1.00} & \textbf{0.92}
& 1.00 & 0.91 \\
Brown & \textbf{0.94} & \textbf{0.47} & \textbf{0.98} & \textbf{0.74} & {1.00}
& \textbf{0.93} & \textbf{1.00} & \textbf{0.97} \\
Noether & 0.75 & 0.33 & 0.90 & 0.67 & 0.95 & 0.74 & 0.97 & 0.87 \\
Adaptive & 0.92 & 0.44 & 0.97 & \textbf{0.74} & 1.00 & 0.90 & \textbf{1.00}
& \textbf{0.97} \\
\hline
\end{tabular*}
\end{table}

In Tables \ref{tabsimhalf} and \ref{tabsimquarter}, there is an additive effect and no bias from unobserved
covariates, that is, $Y_{i}=\tau+\varepsilon_{i}$ and the $\varepsilon
_{i}$ are
independent and identically distributed with a Normal, a logistic or a central
$t$-distribution with 3 degrees of freedom. In Table \ref{tabsimhalf}, the effect is half
the standard deviation $\sigma$ of the $\varepsilon_{i}$'s, $\tau/\sigma=1/2$,
whereas in Table \ref{tabsimquarter}, the effect is either $\tau/\sigma=1/4$ or $\tau
/\sigma=3/4$.

Each sampling situation is replicated 10,000 times. Therefore, the standard
error of the simulated power is at most $\sqrt{1/( 4\times10\mbox{,}000)
}=0.005$. In each sampling situation for each $\Gamma$, the two highest
powers are in \textit{bold}.

Based on Table \ref{tabeff}, we expect Wilcoxon's statistic to have the highest
power for
$\Gamma=1$. Based on Table \ref{tabds}, we expect that for sufficiently large
$\Gamma$
and $I$, Noether's statistic will have the highest power. Combining
Tables \ref{tabeff}
and \ref{tabds}, we see that, for the $t$-distribution with 3 degrees of freedom,
Brown's statistic is much more efficient than Noether's statistic but
has only
slightly inferior design sensitivity, so Brown's statistic could have higher
power for quite large $I$. Proposition \ref{PropDS} suggests that the
adaptive procedure has fulfilled its potential if it has power close to the
maximum of the powers of Brown's and Noether's statistics. With a few
exceptions, these expectations are confirmed in Tables \ref{tabsimhalf} and \ref{tabsimquarter}.
Notably, in
Tables \ref{tabsimhalf} and \ref{tabsimquarter}, the adaptive procedure is never very bad, whereas other
statistics perform poorly in some cases; for instance, in Table \ref{tabsimquarter} the power
loss is 90\% for Wilcoxon's statistic compared to Noether's statistic for
$I=500$ pairs, Normal errors, $\tau/\sigma=3/4$.%

It is useful to contrast Tables \ref{tabds}, \ref{tabsimhalf} and \ref{tabsimquarter}. For instance, the
design sensitivity (as $I\rightarrow\infty$) of Noether's statistic is
$\widetilde{\Gamma}=4.97$ for matched pair differences~$Y_{i}$ that are
$Y_{i}\sim_{\mathrm{i.i.d.}}N( \frac{1}{2},1) $ in Table \ref{tabds},
but the power is only 15\% in this case at
$\Gamma=4<4.97=\widetilde{\Gamma}$ for $I=250$ pairs in Table \ref{tabsimhalf}. That
is, if $Y_{i}\sim_{\mathrm{i.i.d.}}N( \frac{1}{2},1) $ with $I=250$
pairs, there is an 89\% chance the results will be sensitive at
$\Gamma=4$, even though as $I\rightarrow\infty$ the same distribution
would eventually be seen to be insensitive at $\Gamma=4$. The design
sensitivity $\widetilde{\Gamma}$ in Table \ref{tabds} refers to the
limit as $I\rightarrow \infty$, so results will typically become
sensitive at a smaller $\Gamma$, $\Gamma<\widetilde{\Gamma}$, in a
finite sample, $I<\infty$. Although Noether's statistic is always best
in Table \ref{tabds}, it is not always best in Tables \ref{tabsimhalf} and \ref{tabsimquarter}; however,
the adaptive test is never far behind the best test in Tables \ref{tabsimhalf} and~\ref{tabsimquarter}.

\subsection{\texorpdfstring{Ties.}{Ties}}
\label{ssTies}

Ties are addressed in a straightforward manner when testing~$H_{\tau}$.
Providing fewer than $( 1-\lambda_{2}) I$ of the $Y_{i}-\tau$
are equal to zero, no adjustment for zero differences is needed in the
discussion in Section \ref{ssExactNulol}; that is, Brown's statistic
and the
adaptive procedure require no adjustment unless more than $1/3$ of the sample
is tied at zero. For ties among the $\vert Y_{i}-\tau\vert$, use
average ranks in computing $q_{i}$; then $I_{1}$ and $I_{2}$ are random
variables that depend upon the pattern of ties, but the procedure in
Section \ref{ssExactNulol} yields a test that is conditionally distribution-free
given the realized values of $I_{1}$ and~$I_{2}$.

\section{\texorpdfstring{Use of the adaptive procedure in the study of treatments for ovarian
cancer.}{Use of the adaptive procedure in the study of treatments for ovarian
cancer}}
\label{secExampleAnalysis}

The matched pair difference in weeks with toxicity in the first year after
diagnosis is highly significant in randomization tests; for instance, the
randomization based $P$-value from Wilcoxon's signed rank test is less than
$10^{-6}$. For $\Gamma=1.3$ and $\Gamma=1.6$, the upper bounds on the
one-sided $P$-value from Wilcoxon's test are, respectively, 0.0032 and 0.128,
so a bias of $\Gamma=1.3$ could not easily produce the observed value of
Wilcoxon's statistic, but a~bias of $\Gamma=1.6$ could do so. In contrast,
the upper bound on the one-sided $P$-value from the adaptive procedure is
0.004 for $\Gamma=1.6$, so the magnitude of bias that would explain the
behavior of Wilcoxon's statistic does not begin to explain the behavior
of the
adaptive test. [The $P$-value at $\Gamma=1.6$ for the adaptive test is the
smallest $\alpha$ in (\ref{eqCrit1})--(\ref{eqCrit3}) that leads to rejection.]
The upper bound on the $P$-value from the adaptive test crosses 0.05 between
$\Gamma=1.96$ to $\Gamma=1.97$. At $\Gamma=1.96$, the adaptive procedure
rejects based on Noether's statistic, which if used on its own would
have an
upper-bound on its one-sided $P$-value of 0.038. To put these
quantities in
context using the approach in \citet{RosSil09}, $\Gamma=2$
corresponds with an unobserved covariate $u_{ij}$ that produces a three-fold
increase in the odds of greater toxicity and a five-fold increase in
the odds
of treatment by a medical oncologist, so the adaptive test reports
considerably less sensitivity to bias from an unmeasured covariate than does
Wilcoxon's test.

Similar results are found over the first five years. The upper bound on the
$P$-value from Wilcoxon's test is 0.080 for $\Gamma=1.7$, whereas for the
adaptive test, the upper bound on the $P$-value is 0.047 for $\Gamma
=2.2 $.
As before, it is Noether's test, not Brown's test, that leads the adaptive
test to reject.

To illustrate the calculations for toxicity in the first year, allowing for
ties, Noether's statistic looks at the largest $I_{1}=100$ of the $\vert
Y_{i}\vert$ finding $B_{1}=82$ of these have $Y_{i}>0$, whereas Brown's
statistic looks at the largest $I_{1}+I_{2}=110+106=226$ of the $\vert
Y_{i}\vert$ and the statistic has value $T=2B_{1}+B_{2}=222$. Using
the binomial distribution as discussed in Section \ref{ssExactNulol} with
$\Gamma=1.96$, one finds $\Pr( \overline{\overline{B}}_{1}\geq82)
=0.0381$, $\Pr( \overline{\overline{T}}_{1}\geq237) =0.0293$,
$\Pr( \overline{\overline{B}}_{1}\geq82 \mbox{ or } \overline
{\overline{T}}_{1}\geq237) =0.0475$, so the adaptive procedure rejects
at the 0.05 level for every bias less than $\Gamma=1.96$, but only Noether's
test, not Brown's test, would have led to rejection used on its own.

In Section \ref{ssDiscVariations} the choice of $( \lambda_{1},\lambda
_{2}) $ is discussed. As a prelude to that discussion, consider the
results of the sensitivity analysis for toxicity in the first year for two
choices of $( \lambda_{1},\lambda_{2}) $ besides $(
1/3,2/3) $. If $\lambda_{1}=1/6$ and $\lambda_{2}=2/6$ are used in
place of $\lambda_{1}=1/3$ and $\lambda_{2}=2/3$, the adaptive
procedure has
an upper bound on the one-sided $P$-value of 0.046 for $\Gamma=3.3$. If
$\lambda_{1}=1/8$ and $\lambda_{2}=2/8$ are used, the adaptive
procedure has
an upper bound on the one-sided $P$-value of 0.046 for $\Gamma=3.7$. Using
$\lambda_{1}=1/8$ and allowing for ties in the $I=344$ pairs, Noether's
statistic focuses on the 45 of 344 pairs with the largest $\vert
Y_{i}\vert$ and finds that 41 of these 45 pairs have $Y_{i}>0$. In
words, when there was a large difference in weeks with toxicity, it was
usually the result of greater toxicity in a patient treated by a medical
oncologist, and this seems unlikely to have occurred by chance if the
magnitude of bias from nonrandom assignment is $\Gamma\leq3.7$.

\section{\texorpdfstring{Discussion.}{Discussion}}
\label{secDisc}

\subsection{\texorpdfstring{Variations on a theme: Other $\lambda$'s; other statistics.}%
{Variations on a theme: Other lambda's; other statistics}}
\label{ssDiscVariations}

The adaptive procedure in Section \ref{secAdaptive} uses two compatible
tests statistics from the statistical literature. Brown's
(\citeyear{Bro81}) statistic was designed to be a serious competitor of
Wilcoxon's statistic in a randomized experiment without bias, yet
Brown's statistic has higher design sensitivity when errors are Normal
or logistic or $t$-distributed with 3 degrees of freedom. The version
of Noether's (\citeyear{Noe73}) test used here has poor Pitman
efficiency in these cases but much better design sensitivity. So the
adaptive procedure adapts between a procedure with good Pitman
efficiency with good design sensitivity and a procedure with poor
Pitman efficiency and excellent design sensitivity. There are, of
course, many possible variations on this theme, some more promising
than others.

The statistics of Brown and Noether take one or two large steps, but otherwise
are constant as functions of the ranks $q_{i}$ of the $\vert
Y_{i}\vert$. Are large flat steps useful? Both statistics decrease
the weight attached to small $\vert Y_{i}\vert$ and increase the
weight attached to large $\vert Y_{i}\vert$ without emphasizing
the extremely large $\vert Y_{i}\vert$. Would a gradual increase
be better than a step? Consider ranks that equal $q_{i}/I$ if $q_{i}%
/I\geq1-\lambda$ and equal 0 if $q_{i}/I<1-\lambda$; call this the
``$( 1-\lambda) $-step Wilcoxon
statistic'' because it uses Wilcoxon's ranks above
$1-\lambda$. Wilcoxon's statistic is the 0-step Wilcoxon statistic. The
$2/3$-step Wilcoxon statistic takes a step where Noether's statistic
takes a
step, but it increases gradually thereafter, and the $1/3$-step Wilcoxon
statistic takes a step where Brown's statistic takes its first step,
but it
increases gradually thereafter. Table \ref{tabcompare} contrasts the design sensitivities
of Brown's statistic, Noether's statistic and comparable step-Wilcoxon
statistics in the case of an additive treatment effect whose magnitude
is half
the standard deviation of the errors. While the difference between Brown's
statistic and Noether's statistic is large, the difference between
either of
these and its comparable step-Wilcoxon statistic is not large.%

%
\begin{table}
\tablewidth=285pt
\caption{Design sensitivities for Brown's statistic, Noether's
statistic and for two comparable step-Wilcoxon statistics. The table
refers to an additive treatment effect that is half the standard
deviation of the errors, for errors with a Normal distribution, a
logistic distribution or a $t$-distribution with 3 degrees of freedom}
\label{tabcompare}
\begin{tabular*}{\tablewidth}{@{\extracolsep{\fill}}l ccc@{}}
\hline
& \textbf{Normal} & \textbf{Logistic} & $\bolds{t}$ \textbf{3 df} \\
\hline
Brown & 3.60 & 3.83 & 5.39 \\
$1/3$-step Wilcoxon & 3.60 & 3.83 & 5.35\\
[4pt]
Noether & 4.97 & 4.72 & 5.77\\
$2/3$-step Wilcoxon & 5.20 & 4.80 & 5.64 \\
\hline
\end{tabular*}
\end{table}

Brown's statistic focuses on the largest $2/3$ of the $\vert Y_{i}%
\vert$, while Noether's statistic focuses on the largest $1/3$ of $\vert
Y_{i}\vert$. In the example in Section \ref{secEG}, further tinkering
with $\lambda_{1}$ and $\lambda_{2}$ led to greater insensitivity to
unmeasured bias. \citet{MarHet82} discuss the choice of~$\lambda_{1}$ and~$\lambda_{2}$ from the perspective of Pitman
efficiency. Table \ref{tabcompare2} compares the design sensitivities
for several values of $\lambda
_{1}$ with $\lambda_{2}=2\lambda_{1}$.

%
\begin{table}[b]
\caption{Design sensitivities for the Brown--Markowski--Hettmansperger
statistic, Noether's statistic and the adaptive statistic for various
values of $\lambda_{1}$ with $\lambda_{2}=2\lambda _{1}$. The table
refers to an additive treatment effect that is half the standard
deviation of the errors, for errors with a Normal distribution, a
logistic distribution or a $t$-distribution with 3 degrees of freedom.
The largest design sensitivity in a sampling situation (or in a column)
is in \textup{bold}}
\label{tabcompare2}
\begin{tabular*}{\tablewidth}{@{\extracolsep{\fill}}l c ccc@{}}
\hline
& $\bolds{\lambda_{1}}$ & \textbf{Normal} & \textbf{Logistic}
& $\bolds{t}$ \textbf{3 df} \\
\hline
Brown--Markowski--Hettmansperger & $1/3$ & 3.60 & 3.83 & 5.39\\
Noether & $1/3$ & 4.97 & 4.72 & \textbf{5.77}\\
Adaptive & $1/3$ & 4.97 & 4.72 & \textbf{5.77}\\
[4pt]
Brown--Markowski--Hettmansperger & $1/4$ & 4.36 & 4.37 & 5.67\\
Noether & $1/4$ & 5.87 & 5.06 & 5.53\\
Adaptive & $1/4$ & 5.87 & 5.06 & 5.67\\
[4pt]
Brown--Markowski--Hettmansperger & $1/6$ & 5.58 & 4.93 & 5.51\\
Noether & $1/6$ & 7.28 & 5.41 & 5.03\\
Adaptive & $1/6$ & 7.28 & 5.41 & 5.51\\
[4pt]
Brown--Markowski--Hettmansperger & $1/8$ & 6.55 & 5.23 & 5.20\\
Noether & $1/8$ & \textbf{8.40} & \textbf{5.59} & 4.64\\
Adaptive & $1/8$ & \textbf{8.40} & \textbf{5.59} & 5.20\\
\hline
\end{tabular*}
\end{table}

In thinking about Table \ref{tabcompare2}, several cautions are needed.
First, the design sensitivity refers to a limit as the number $I$ of
pairs increases $I\rightarrow\infty$, so Table \ref{tabcompare2} is
unlikely to offer useful guidance unless $I\lambda_{1}$ is a reasonably
large number. With $I=100$ pairs and $\lambda_{1}=1/8$, there are only
13 pairs counted in Noether's statistic, so asymptotic theory is not
likely to provide useful guidance. Second, the Pitman efficiencies for
Noether's statistic with $\lambda_{1}<1/3$ are substantially worse than
the already disappointing values shown in Table \ref{tabeff}, so Table
\ref{tabcompare2} is only relevant when the sample size $I$ is so large
that the design sensitivity has come to dominate the Pitman efficiency,
as it will do in the limit as $I\rightarrow\infty$ because the power
function tends to a step function dropping from power 1 to power 0 at
$\widetilde{\Gamma}$. There are, however, many large observational
studies, for example, \citet{Voletal07} conducted an observational
study of 8.5 million hospital admissions. Third, the columns of Table
\ref{tabcompare2} refer to distributions that differ greatly in their
tails, so an answer that depends strongly upon which column is
considered is an answer that depends strongly on the behavior of the
most extreme observations.

With these cautions firmly in mind, consider Table \ref{tabcompare2}. In Table \ref{tabcompare2},
$\lambda_{1}=1/8$ is best for the Normal and logistic distributions and
$\lambda_{1}=1/3$ is best for the $t$-distribution with 3 degrees of freedom;
see Rosenbaum [(\citeyear{Ros10N1}), Figure 2] for a heuristic explanation of the
relationship between tail behavior, weights and sensitivity to bias. With
smaller $\lambda_{1}$'s, the adaptive procedure looks attractive: for
$\lambda_{1}=1/8$ it uses Noether's test to advantage for Normal errors
and it
uses the Brown--Markowski--Hettmansperger test for $t$-errors. Notably, in
Table \ref{tabcompare2}, the adaptive procedure exhibits relatively stable performance
as~$\lambda_{1}$ decreases for the $t$-distribution, but it captures large gains
for the Normal and logistic distributions.

\subsection{\texorpdfstring{Are large observational studies less susceptible to unmeasured
biases\textup{?}}{Are large observational studies less susceptible to unmeasured
biases}}
\label{ssDiscAreLarge}

Section \ref{secIntro} began with the question: are large observational
studies less susceptible to unmeasured biases? The success of the adaptive
procedure suggests that this question is incorrectly posed. An observational
study is sensitive to biases of a certain magnitude, and the sample
size is
not the key element in determining this. However, a poor choice of test
statistic---perhaps the Wilcoxon statistic---may lead to a sensitivity
analysis that exaggerates the degree of sensitivity to unmeasured
biases. A
good choice of test statistic may depend upon features of the observable
distributions that are unknown to the investigator prior to the investigation.
To the extent that a large sample size permits us to see clearly these
features of observable distributions, it may let us adapt the statistical
analysis so that a poor choice of test statistic does not exaggerate the
degree of sensitivity to unmeasured biases.


%

\printaddresses


\begin{thebibliography}{52}

\bibitem[\protect\citeauthoryear{Brown}{1981}]{Bro81}
\begin{barticle}[mr]
\bauthor{\bsnm{Brown},~\bfnm{B.~M.}\binits{B.~M.}}
(\byear{1981}).
\btitle{Symmetric quantile averages and related estimators}.
\bjournal{Biometrika}
\bvolume{68}
\bpages{235--242}.
\bid{doi={10.1093/biomet/68.1.235}, issn={0006-3444}, mr={0614960}}
\bptok{imsref}%
\end{barticle}
\endbibitem

\bibitem[\protect\citeauthoryear{Cochran}{1965}]{Coc}
\begin{barticle}[auto:STB|2011/12/02|17:21:01]
\bauthor{\bsnm{Cochran},~\bfnm{W.~G.}\binits{W.~G.}}
(\byear{1965}).
\btitle{The planning of observational studies of human populations (with
  discussion)}.
\bjournal{J. Roy. Statist. Soc. Ser. A}
\bvolume{128}
\bpages{234--266}.
\bptok{imsref}%
\end{barticle}
\endbibitem

\bibitem[\protect\citeauthoryear{Copas and Eguchi}{2001}]{CopEgu01}
\begin{barticle}[mr]
\bauthor{\bsnm{Copas},~\bfnm{John}\binits{J.}} \AND
  \bauthor{\bsnm{Eguchi},~\bfnm{Shinto}\binits{S.}}
(\byear{2001}).
\btitle{Local sensitivity approximations for selectivity bias}.
\bjournal{J.~R.~Stat. Soc. Ser. B Stat. Methodol.}
\bvolume{63}
\bpages{871--895}.
\bid{doi={10.1111/1467-9868.00318}, issn={1369-7412}, mr={1872072}}
\bptok{imsref}%
\end{barticle}
\endbibitem

\bibitem[\protect\citeauthoryear{Cornfield et~al.}{1959}]{CORetal59}
\begin{barticle}[pbm]
\bauthor{\bsnm{Cornfield},~\bfnm{J.}\binits{J.}},
  \bauthor{\bsnm{Haenszel},~\bfnm{W.}\binits{W.}},
  \bauthor{\bsnm{Hammond},~\bfnm{E.~C.}\binits{E.~C.}},
  \bauthor{\bsnm{Lilienfeld},~\bfnm{A.~M.}\binits{A.~M.}},
  \bauthor{\bsnm{Shimkin},~\bfnm{M.~B.}\binits{M.~B.}} \AND
  \bauthor{\bsnm{Wynder},~\bfnm{E.~L.}\binits{E.~L.}}
(\byear{1959}).
\btitle{Smoking and lung cancer}. 
\bjournal{J. Natl. Cancer Inst.}
\bvolume{22}
\bpages{173--203}.
\bid{issn={0027-8874}, pmid={13621204}}
\bptnote{check related}%
\bptok{imsref}%
\end{barticle}
\endbibitem

\bibitem[\protect\citeauthoryear{Cox}{1975}]{Cox75}
\begin{barticle}[mr]
\bauthor{\bsnm{Cox},~\bfnm{D.~R.}\binits{D.~R.}}
(\byear{1975}).
\btitle{A note on data-splitting for the evaluation of significance levels}.
\bjournal{Biometrika}
\bvolume{62}
\bpages{441--444}.
\bid{issn={0006-3444}, mr={0378189}}
\bptok{imsref}%
\end{barticle}
\endbibitem

\bibitem[\protect\citeauthoryear{Diprete and Gangl}{2004}]{DipGan04}
\begin{barticle}[auto:STB|2011/12/02|17:21:01]
\bauthor{\bsnm{Diprete},~\bfnm{T.~A.}\binits{T.~A.}} \AND
  \bauthor{\bsnm{Gangl},~\bfnm{M.}\binits{M.}}
(\byear{2004}).
\btitle{Assessing bias in the estimation of causal effects}.
\bjournal{Sociol. Method.}
\bvolume{34}
\bpages{271--310}.
\bptok{imsref}%
\end{barticle}
\endbibitem

\bibitem[\protect\citeauthoryear{Egleston, Scharfstein and
  MacKenzie}{2009}]{EglSchMac09}
\begin{barticle}[mr]
\bauthor{\bsnm{Egleston},~\bfnm{Brian~L.}\binits{B.~L.}},
  \bauthor{\bsnm{Scharfstein},~\bfnm{Daniel~O.}\binits{D.~O.}} \AND
  \bauthor{\bsnm{MacKenzie},~\bfnm{Ellen}\binits{E.}}
(\byear{2009}).
\btitle{On estimation of the survivor average causal effect in observational
  studies when important confounders are missing due to death}.
\bjournal{Biometrics}
\bvolume{65}
\bpages{497--504}.
\bid{doi={10.1111/j.1541-0420.2008.01111.x}, issn={0006-341X}, mr={2751473}}
\bptok{imsref}%
\end{barticle}
\endbibitem

\bibitem[\protect\citeauthoryear{Fisher}{1935}]{Fis35}
\begin{bbook}[auto:STB|2011/12/02|17:21:01]
\bauthor{\bsnm{Fisher},~\bfnm{R.~A.}\binits{R.~A.}}
(\byear{1935}).
\btitle{Design of Experiments}.
\bpublisher{Oliver \& Boyd}, \baddress{Edinburgh}.
\bptok{imsref}%
\end{bbook}
\endbibitem

\bibitem[\protect\citeauthoryear{Frangakis and Rubin}{1999}]{FraRub99}
\begin{barticle}[mr]
\bauthor{\bsnm{Frangakis},~\bfnm{Constantine~E.}\binits{C.~E.}} \AND
  \bauthor{\bsnm{Rubin},~\bfnm{Donald~B.}\binits{D.~B.}}
(\byear{1999}).
\btitle{Addressing complications of intention-to-treat analysis in the combined
  presence of all-or-none treatment-noncompliance and subsequent missing
  outcomes}.
\bjournal{Biometrika}
\bvolume{86}
\bpages{365--379}.
\bid{doi={10.1093/biomet/86.2.365}, issn={0006-3444}, mr={1705410}}
\bptnote{check year}%
\bptok{imsref}%
\end{barticle}
\endbibitem

\bibitem[\protect\citeauthoryear{Gadbury}{2001}]{Gad01}
\begin{barticle}[mr]
\bauthor{\bsnm{Gadbury},~\bfnm{Gary~L.}\binits{G.~L.}}
(\byear{2001}).
\btitle{Randomization inference and bias of standard errors}.
\bjournal{Amer. Statist.}
\bvolume{55}
\bpages{310--313}.
\bid{doi={10.1198/000313001753272268}, issn={0003-1305}, mr={1939365}}
\bptok{imsref}%
\end{barticle}
\endbibitem

\bibitem[\protect\citeauthoryear{Gastwirth}{1966}]{Gas66}
\begin{barticle}[mr]
\bauthor{\bsnm{Gastwirth},~\bfnm{Joseph~L.}\binits{J.~L.}}
(\byear{1966}).
\btitle{On robust procedures}.
\bjournal{J. Amer. Statist. Assoc.}
\bvolume{61}
\bpages{929--948}.
\bid{issn={0162-1459}, mr={0205397}}
\bptnote{check year}%
\bptok{imsref}%
\end{barticle}
\endbibitem

\bibitem[\protect\citeauthoryear{Gastwirth}{1992}]{Gas92}
\begin{barticle}[auto:STB|2011/12/02|17:21:01]
\bauthor{\bsnm{Gastwirth},~\bfnm{J.~L.}\binits{J.~L.}}
(\byear{1992}).
\btitle{Methods for assessing the sensitivity of statistical comparisons used
  in Title VII cases to omitted variables}.
\bjournal{Jurimetrics}
\bvolume{33}
\bpages{19--34}.
\bptok{imsref}%
\end{barticle}
\endbibitem

\bibitem[\protect\citeauthoryear{Gilbert, Bosch and
  Hudgens}{2003}]{GilBosHud03}
\begin{barticle}[mr]
\bauthor{\bsnm{Gilbert},~\bfnm{Peter~B.}\binits{P.~B.}},
  \bauthor{\bsnm{Bosch},~\bfnm{Ronald~J.}\binits{R.~J.}} \AND
  \bauthor{\bsnm{Hudgens},~\bfnm{Michael~G.}\binits{M.~G.}}
(\byear{2003}).
\btitle{Sensitivity analysis for the assessment of causal vaccine effects on
  viral load in {HIV} vaccine trials}.
\bjournal{Biometrics}
\bvolume{59}
\bpages{531--541}.
\bid{doi={10.1111/1541-0420.00063}, issn={0006-341X}, mr={2004258}}
\bptok{imsref}%
\end{barticle}
\endbibitem

\bibitem[\protect\citeauthoryear{Groeneveld}{1972}]{Groeneveld72}
\begin{barticle}[mr]
\bauthor{\bsnm{Groeneveld},~\bfnm{R.~A.}\binits{R.~A.}}
(\byear{1972}).
\btitle{Asymptotically optimal group rank tests for location}.
\bjournal{J.~Amer. Statist. Assoc.}
\bvolume{67}
\bpages{847--849}.
\bptok{imsref}%
\end{barticle}
\endbibitem

\bibitem[\protect\citeauthoryear{Heller, Rosenbaum and
  Small}{2009}]{HelRosSma09}
\begin{barticle}[mr]
\bauthor{\bsnm{Heller},~\bfnm{Ruth}\binits{R.}},
  \bauthor{\bsnm{Rosenbaum},~\bfnm{Paul~R.}\binits{P.~R.}} \AND
  \bauthor{\bsnm{Small},~\bfnm{Dylan~S.}\binits{D.~S.}}
(\byear{2009}).
\btitle{Split samples and design sensitivity in observational studies}.
\bjournal{J. Amer. Statist. Assoc.}
\bvolume{104}
\bpages{1090--1101}.
\bid{doi={10.1198/jasa.2009.tm08338}, issn={0162-1459}, mr={2750238}}
\bptok{imsref}%
\end{barticle}
\endbibitem

\bibitem[\protect\citeauthoryear{Hodges and Lehmann}{1963}]{HodLeh63}
\begin{barticle}[mr]
\bauthor{\bsnm{Hodges},~\bfnm{J.~L.}\binits{J.~L.} \bsuffix{Jr.}} \AND
  \bauthor{\bsnm{Lehmann},~\bfnm{E.~L.}\binits{E.~L.}}
(\byear{1963}).
\btitle{Estimates of location based on rank tests}.
\bjournal{Ann. Math. Statist.}
\bvolume{34}
\bpages{598--611}.
\bid{issn={0003-4851}, mr={0152070}}
\bptok{imsref}%
\end{barticle}
\endbibitem

\bibitem[\protect\citeauthoryear{Hogg}{1974}]{Hog74}
\begin{barticle}[mr]
\bauthor{\bsnm{Hogg},~\bfnm{Robert~V.}\binits{R.~V.}}
(\byear{1974}).
\btitle{Adaptive robust procedures: A partial review and some suggestions for
  future applications and theory (with discussion)}.
\bjournal{J. Amer. Statist. Assoc.}
\bvolume{69}
\bpages{909--923}.
\bptnote{check related}%
\bptok{imsref}%
\end{barticle}
\endbibitem

\bibitem[\protect\citeauthoryear{Hosman, Hansen and
  Holland}{2010}]{HosHanHol10}
\begin{barticle}[mr]
\bauthor{\bsnm{Hosman},~\bfnm{Carrie~A.}\binits{C.~A.}},
  \bauthor{\bsnm{Hansen},~\bfnm{Ben~B.}\binits{B.~B.}} \AND
  \bauthor{\bsnm{Holland},~\bfnm{Paul~W.}\binits{P.~W.}}
(\byear{2010}).
\btitle{The sensitivity of linear regression coefficients' confidence limits to
  the omission of a confounder}.
\bjournal{Ann. Appl. Stat.}
\bvolume{4}
\bpages{849--870}.
\bid{doi={10.1214/09-AOAS315}, issn={1932-6157}, mr={2758424}}
\bptok{imsref}%
\end{barticle}
\endbibitem

\bibitem[\protect\citeauthoryear{Imbens}{2003}]{Imb03}
\begin{barticle}[auto:STB|2011/12/02|17:21:01]
\bauthor{\bsnm{Imbens},~\bfnm{G.~W.}\binits{G.~W.}}
(\byear{2003}).
\btitle{Sensitivity to exogeneity assumptions in program evaluation}.
\bjournal{Am. Econ. Rev.}
\bvolume{93}
\bpages{126--132}.
\bptok{imsref}%
\end{barticle}
\endbibitem

\bibitem[\protect\citeauthoryear{Jones}{1979}]{Jon79}
\begin{barticle}[mr]
\bauthor{\bsnm{Jones},~\bfnm{Douglas~H.}\binits{D.~H.}}
(\byear{1979}).
\btitle{An efficient adaptive distribution-free test for location}.
\bjournal{J. Amer. Statist. Assoc.}
\bvolume{74}
\bpages{822--828}.
\bid{issn={0003-1291}, mr={0556475}}
\bptok{imsref}%
\end{barticle}
\endbibitem

\bibitem[\protect\citeauthoryear{Lehmann}{1975}]{Leh75}
\begin{bbook}[auto:STB|2011/12/02|17:21:01]
\bauthor{\bsnm{Lehmann},~\bfnm{E.~L.}\binits{E.~L.}}
(\byear{1975}).
\btitle{Nonparametrics}.
\bpublisher{Holden Day}, \baddress{San Francisco}.
\bptok{imsref}%
\end{bbook}
\endbibitem

\bibitem[\protect\citeauthoryear{Lin, Psaty and Kronmal}{1998}]{LinPsaKro98}
\begin{barticle}[pbm]
\bauthor{\bsnm{Lin},~\bfnm{D.~Y.}\binits{D.~Y.}},
  \bauthor{\bsnm{Psaty},~\bfnm{B.~M.}\binits{B.~M.}} \AND
  \bauthor{\bsnm{Kronmal},~\bfnm{R.~A.}\binits{R.~A.}}
(\byear{1998}).
\btitle{Assessing the sensitivity of regression results to unmeasured
  confounders in observational studies}.
\bjournal{Biometrics}
\bvolume{54}
\bpages{948--963}.
\bid{issn={0006-341X}, pmid={9750244}}
\bptok{imsref}%
\end{barticle}
\endbibitem

\bibitem[\protect\citeauthoryear{Marcus}{1997}]{Mar97}
\begin{barticle}[auto:STB|2011/12/02|17:21:01]
\bauthor{\bsnm{Marcus},~\bfnm{S.~M.}\binits{S.~M.}}
(\byear{1997}).
\btitle{Using omitted variable bias to assess uncertainty in the estimation of
  an AIDS education treatment effect}.
\bjournal{J. Educ. Behav. Statist.}
\bvolume{22}
\bpages{193--201}.
\bptok{imsref}%
\end{barticle}
\endbibitem

\bibitem[\protect\citeauthoryear{Maritz}{1979}]{Mar79}
\begin{barticle}[mr]
\bauthor{\bsnm{Maritz},~\bfnm{J.~S.}\binits{J.~S.}}
(\byear{1979}).
\btitle{A note on exact robust confidence intervals for location}.
\bjournal{Biometrika}
\bvolume{66}
\bpages{163--166}.
\bid{doi={10.1093/biomet/66.1.163}, issn={0006-3444}, mr={0529161}}
\bptok{imsref}%
\end{barticle}
\endbibitem

\bibitem[\protect\citeauthoryear{Markowski and Hettmansperger}{1982}]{MarHet82}
\begin{barticle}[mr]
\bauthor{\bsnm{Markowski},~\bfnm{Edward~P.}\binits{E.~P.}} \AND
  \bauthor{\bsnm{Hettmansperger},~\bfnm{Thomas~P.}\binits{T.~P.}}
(\byear{1982}).
\btitle{Inference based on simple rank step score statistics for the location
  model}.
\bjournal{J. Amer. Statist. Assoc.}
\bvolume{77}
\bpages{901--907}.
\bid{issn={0162-1459}, mr={0686416}}
\bptok{imsref}%
\end{barticle}
\endbibitem

\bibitem[\protect\citeauthoryear{McCandless, Gustafson and
  Levy}{2007}]{McCGusLev07}
\begin{barticle}[mr]
\bauthor{\bsnm{McCandless},~\bfnm{Lawrence~C.}\binits{L.~C.}},
  \bauthor{\bsnm{Gustafson},~\bfnm{Paul}\binits{P.}} \AND
  \bauthor{\bsnm{Levy},~\bfnm{Adrian}\binits{A.}}
(\byear{2007}).
\btitle{Bayesian sensitivity analysis for unmeasured confounding in
  observational studies}.
\bjournal{Stat. Med.}
\bvolume{26}
\bpages{2331--2347}.
\bid{doi={10.1002/sim.2711}, issn={0277-6715}, mr={2368419}}
\bptok{imsref}%
\end{barticle}
\endbibitem

\bibitem[\protect\citeauthoryear{Neyman}{1923}]{Neyman23}
\begin{barticle}[mr]
\bauthor{\bsnm{Neyman},~\bfnm{J.}\binits{J.}}
(\byear{1923}).
\btitle{On the application of probability theory to agricultural experiments}.
\bjournal{Statist. Sci.}
\bvolume{5}
\bpages{463--480}.
\bptok{imsref}%
\end{barticle}
\endbibitem

\bibitem[\protect\citeauthoryear{Noether}{1973}]{Noe73}
\begin{barticle}[auto:STB|2011/12/02|17:21:01]
\bauthor{\bsnm{Noether},~\bfnm{G.}\binits{G.}}
(\byear{1973}).
\btitle{Some distribution-free confidence intervals for the center of a
  symmetric distribution}.
\bjournal{J. Amer. Statist. Assoc.}
\bvolume{68}
\bpages{716--719}.
\bptok{imsref}%
\end{barticle}
\endbibitem

\bibitem[\protect\citeauthoryear{Policello and Hettmansperger}{1976}]{PolHet76}
\begin{barticle}[auto:STB|2011/12/02|17:21:01]
\bauthor{\bsnm{Policello},~\bfnm{G.~E.}\binits{G.~E.}} \AND
  \bauthor{\bsnm{Hettmansperger},~\bfnm{T.~P.}\binits{T.~P.}}
(\byear{1976}).
\btitle{Adaptive robust procedures for the one-sample location problem}.
\bjournal{J. Amer. Statist. Assoc.}
\bvolume{71}
\bpages{624--633}.
\bptok{imsref}%
\end{barticle}
\endbibitem

\bibitem[\protect\citeauthoryear{Reiter}{2000}]{Rei00}
\begin{barticle}[mr]
\bauthor{\bsnm{Reiter},~\bfnm{Jerome}\binits{J.}}
(\byear{2000}).
\btitle{Using {s}tatistics to {d}etermine {c}ausal {r}elationships}.
\bjournal{Amer. Math. Monthly}
\bvolume{107}
\bpages{24--32}.
\bid{doi={10.2307/2589374}, issn={0002-9890}, mr={1543589}}
\bptok{imsref}%
\end{barticle}
\endbibitem

\bibitem[\protect\citeauthoryear{Rosenbaum}{1993}]{Ros93}
\begin{barticle}[mr]
\bauthor{\bsnm{Rosenbaum},~\bfnm{Paul~R.}\binits{P.~R.}}
(\byear{1993}).
\btitle{Hodges--{L}ehmann point estimates of treatment effect in observational
  studies}.
\bjournal{J. Amer. Statist. Assoc.}
\bvolume{88}
\bpages{1250--1253}.
\bid{issn={0162-1459}, mr={1245357}}
\bptok{imsref}%
\end{barticle}
\endbibitem

\bibitem[\protect\citeauthoryear{Rosenbaum}{2002a}]{Ros02a}
\begin{bbook}[mr]
\bauthor{\bsnm{Rosenbaum},~\bfnm{Paul~R.}\binits{P.~R.}}
(\byear{2002}a).
\btitle{Observational Studies}, \bedition{2nd} ed.
\bpublisher{Springer}, \baddress{New York}.
\bptok{imsref}%
\end{bbook}
\endbibitem

\bibitem[\protect\citeauthoryear{Rosenbaum}{2002b}]{Ros02b}
\begin{barticle}[mr]
\bauthor{\bsnm{Rosenbaum},~\bfnm{Paul~R.}\binits{P.~R.}}
(\byear{2002}b).
\btitle{Covariance adjustment in randomized experiments and observational
  studies}.
\bjournal{Statist. Sci.}
\bvolume{17}
\bpages{286--327}.
\bid{doi={10.1214/ss/1042727942}, issn={0883-4237}, mr={1962487}}
\bptnote{check related}%
\bptok{imsref}%
\end{barticle}
\endbibitem

\bibitem[\protect\citeauthoryear{Rosenbaum}{2004}]{Ros04}
\begin{barticle}[mr]
\bauthor{\bsnm{Rosenbaum},~\bfnm{Paul~R.}\binits{P.~R.}}
(\byear{2004}).
\btitle{Design sensitivity in observational studies}.
\bjournal{Biometrika}
\bvolume{91}
\bpages{153--164}.
\bid{doi={10.1093/biomet/91.1.153}, issn={0006-3444}, mr={2050466}}
\bptok{imsref}%
\end{barticle}
\endbibitem

\bibitem[\protect\citeauthoryear{Rosenbaum}{2005}]{Ros05}
\begin{barticle}[mr]
\bauthor{\bsnm{Rosenbaum},~\bfnm{Paul~R.}\binits{P.~R.}}
(\byear{2005}).
\btitle{Heterogeneity and causality: Unit heterogeneity and design sensitivity
  in observational studies}.
\bjournal{Amer. Statist.}
\bvolume{59}
\bpages{147--152}.
\bid{doi={10.1198/000313005X42831}, issn={0003-1305}, mr={2133562}}
\bptok{imsref}%
\end{barticle}
\endbibitem

\bibitem[\protect\citeauthoryear{Rosenbaum}{2010a}]{Ros10N1}
\begin{barticle}[mr]
\bauthor{\bsnm{Rosenbaum},~\bfnm{Paul~R.}\binits{P.~R.}}
(\byear{2010}a).
\btitle{Design sensitivity and efficiency in observational studies}.
\bjournal{J.~Amer. Statist. Assoc.}
\bvolume{105}
\bpages{692--702}.
\bid{doi={10.1198/jasa.2010.tm09570}, issn={0162-1459}, mr={2724853}}
\bptok{imsref}%
\end{barticle}
\endbibitem

\bibitem[\protect\citeauthoryear{Rosenbaum}{2010b}]{Ros10N2}
\begin{bbook}[mr]
\bauthor{\bsnm{Rosenbaum},~\bfnm{Paul~R.}\binits{P.~R.}}
(\byear{2010}b).
\btitle{Design of Observational Studies}.
\bpublisher{Springer}, \baddress{New York}.
\bid{doi={10.1007/978-1-4419-1213-8}, mr={2561612}}
\bptok{imsref}%
\end{bbook}
\endbibitem

\bibitem[\protect\citeauthoryear{Rosenbaum}{2011}]{Ros}
\begin{barticle}[auto:STB|2011/12/02|17:21:01]
\bauthor{\bsnm{Rosenbaum},~\bfnm{P.~R.}\binits{P.~R.}}
(\byear{2011}).
\btitle{A new u-statistic with superior design sensitivity in
  observational studies}.
\bjournal{Biometrics}
\bvolume{67}
\bpages{1017--1027}.
\bptok{imsref}%
\end{barticle}
\endbibitem

\bibitem[\protect\citeauthoryear{Rosenbaum, Ross and
  Silber}{2007}]{RosRosSil07N1}
\begin{barticle}[mr]
\bauthor{\bsnm{Rosenbaum},~\bfnm{Paul~R.}\binits{P.~R.}},
  \bauthor{\bsnm{Ross},~\bfnm{Richard~N.}\binits{R.~N.}} \AND
  \bauthor{\bsnm{Silber},~\bfnm{Jeffrey~H.}\binits{J.~H.}}
(\byear{2007}).
\btitle{Minimum distance matched sampling with fine balance in an observational
  study of treatment for ovarian cancer}.
\bjournal{J. Amer. Statist. Assoc.}
\bvolume{102}
\bpages{75--83}.
\bid{doi={10.1198/016214506000001059}, issn={0162-1459}, mr={2345534}}
\bptok{imsref}%
\end{barticle}
\endbibitem

\bibitem[\protect\citeauthoryear{Rosenbaum and Rubin}{1983}]{RosRub}
\begin{barticle}[auto:STB|2011/12/02|17:21:01]
\bauthor{\bsnm{Rosenbaum},~\bfnm{P.~R.}\binits{P.~R.}} \AND
  \bauthor{\bsnm{Rubin},~\bfnm{D.~B.}\binits{D.~B.}}
(\byear{1983}).
\btitle{Assessing sensitivity to an unobserved binary covariate in an
  observational study with binary outcome}.
\bjournal{J. R. Stat. Soc. Ser. B Stat. Methodol.}
\bvolume{45}
\bpages{212--218}.
\bptok{imsref}%
\end{barticle}
\endbibitem

\bibitem[\protect\citeauthoryear{Rosenbaum and Silber}{2009}]{RosSil09}
\begin{barticle}[mr]
\bauthor{\bsnm{Rosenbaum},~\bfnm{Paul~R.}\binits{P.~R.}} \AND
  \bauthor{\bsnm{Silber},~\bfnm{Jeffrey~H.}\binits{J.~H.}}
(\byear{2009}).
\btitle{Amplification of sensitivity analysis in matched observational
  studies}.
\bjournal{J. Amer. Statist. Assoc.}
\bvolume{104}
\bpages{1398--1405}.
\bid{doi={10.1198/jasa.2009.tm08470}, issn={0162-1459}, mr={2750570}}
\bptok{imsref}%
\end{barticle}
\endbibitem

\bibitem[\protect\citeauthoryear{Rubin}{1974}]{Rub74}
\begin{barticle}[auto:STB|2011/12/02|17:21:01]
\bauthor{\bsnm{Rubin},~\bfnm{D.~B.}\binits{D.~B.}}
(\byear{1974}).
\btitle{Estimating causal effects of treatments in randomized and nonrandomized
  studies}.
\bjournal{J. Educ. Psych.}
\bvolume{66}
\bpages{688--701}.
\bptok{imsref}%
\end{barticle}
\endbibitem

\bibitem[\protect\citeauthoryear{Rubin}{1979}]{Rub79}
\begin{barticle}[auto:STB|2011/12/02|17:21:01]
\bauthor{\bsnm{Rubin},~\bfnm{D.~B.}\binits{D.~B.}}
(\byear{1979}).
\btitle{Using multivariate matched sampling and regression adjustment to
  control bias in observational studies}.
\bjournal{J. Amer. Statist. Assoc.}
\bvolume{74}
\bpages{318--328}.
\bptok{imsref}%
\end{barticle}
\endbibitem

\bibitem[\protect\citeauthoryear{Silber et~al.}{2007}]{RosRosSil07N2}
\begin{barticle}[mr]
\bauthor{\bsnm{Silber},~\bfnm{Jeffrey~H.}\binits{J.~H.}},
  \bauthor{\bsnm{Rosenbaum},~\bfnm{Paul~R.}\binits{P.~R.}},
  \bauthor{\bsnm{Polsky},~\bfnm{D.}\binits{D.}},
  \bauthor{\bsnm{Ross},~\bfnm{Richard~N.}\binits{R.~N.}},
  \bauthor{\bsnm{Even-Shoshan},~\bfnm{O.}\binits{O.}},
  \bauthor{\bsnm{Schwartz},~\bfnm{S.}\binits{S.}},
  \bauthor{\bsnm{Armstrong},~\bfnm{K.~A.}\binits{K.~A.}} \AND
  \bauthor{\bsnm{Randall},~\bfnm{T.~C.}\binits{T.~C.}}
(\byear{2007}).
\btitle{Does ovarian cancer treatment and survival differ by the specialty
  providing chemotherapy?}
\bjournal{J. Clin. Oncol.}
\bvolume{25}
\bpages{1169--1175}.
\bnote{Related editorial: \textbf{25} 1157--1158. Related letters and
  rejoinders: \textbf{25} 3551--3558.}
\bptok{imsref}%
\end{barticle}
\endbibitem

\bibitem[\protect\citeauthoryear{Small}{2007}]{Sma07}
\begin{barticle}[mr]
\bauthor{\bsnm{Small},~\bfnm{Dylan~S.}\binits{D.~S.}}
(\byear{2007}).
\btitle{Sensitivity analysis for instrumental variables regression with
  overidentifying restrictions}.
\bjournal{J. Amer. Statist. Assoc.}
\bvolume{102}
\bpages{1049--1058}.
\bid{doi={10.1198/016214507000000608}, issn={0162-1459}, mr={2411664}}
\bptok{imsref}%
\end{barticle}
\endbibitem

\bibitem[\protect\citeauthoryear{Volpp et~al.}{2007}]{Voletal07}
\begin{barticle}[auto:STB|2011/12/02|17:21:01]
\bauthor{\bsnm{Volpp},~\bfnm{K.~G.}\binits{K.~G.}},
  \bauthor{\bsnm{Rosen},~\bfnm{A.~K.}\binits{A.~K.}},
  \bauthor{\bsnm{Rosenbaum},~\bfnm{P.~R.}\binits{P.~R.}},
  \bauthor{\bsnm{Romano},~\bfnm{P.~S.}\binits{P.~S.}},
  \bauthor{\bsnm{Even-Shoshan},~\bfnm{O.}\binits{O.}},
  \bauthor{\bsnm{Wang},~\bfnm{Y.}\binits{Y.}},
  \bauthor{\bsnm{Bellini},~\bfnm{L.}\binits{L.}},
  \bauthor{\bsnm{Behringer},~\bfnm{T.}\binits{T.}} \AND
  \bauthor{\bsnm{Silber},~\bfnm{J.~H.}\binits{J.~H.}}
(\byear{2007}).
\btitle{Mortality among hospitalized Medicare beneficiaries in the first 2
  years following ACGME resident duty hour reform}.
\bjournal{J. Am. Med. Assoc.}
\bvolume{298}
\bpages{975--983}.
\bptok{imsref}%
\end{barticle}
\endbibitem

\bibitem[\protect\citeauthoryear{Wang and Krieger}{2006}]{WanKri06}
\begin{barticle}[mr]
\bauthor{\bsnm{Wang},~\bfnm{Liansheng}\binits{L.}} \AND
  \bauthor{\bsnm{Krieger},~\bfnm{Abba~M.}\binits{A.~M.}}
(\byear{2006}).
\btitle{Causal conclusions are most sensitive to unobserved binary covariates}.
\bjournal{Stat. Med.}
\bvolume{25}
\bpages{2257--2271}.
\bid{doi={10.1002/sim.2344}, issn={0277-6715}, mr={2240099}}
\bptok{imsref}%
\end{barticle}
\endbibitem

\bibitem[\protect\citeauthoryear{Welch}{1937}]{Wel37}
\begin{barticle}[auto:STB|2011/12/02|17:21:01]
\bauthor{\bsnm{Welch},~\bfnm{B.~L.}\binits{B.~L.}}
(\byear{1937}).
\btitle{On the z-test in randomized blocks and Latin squares}.
\bjournal{Biometrika}
\bvolume{29}
\bpages{21--52}.
\bptok{imsref}%
\end{barticle}
\endbibitem

\bibitem[\protect\citeauthoryear{Wilk and Gnanadesikan}{1968}]{WilGna68}
\begin{barticle}[pbm]
\bauthor{\bsnm{Wilk},~\bfnm{M.~B.}\binits{M.~B.}} \AND
  \bauthor{\bsnm{Gnanadesikan},~\bfnm{R.}\binits{R.}}
(\byear{1968}).
\btitle{Probability plotting methods for the analysis of data}.
\bjournal{Biometrika}
\bvolume{55}
\bpages{1--17}.
\bid{issn={0006-3444}, pmid={5661047}}
\bptok{imsref}%
\end{barticle}
\endbibitem

\bibitem[\protect\citeauthoryear{Wolfe}{1974}]{Wol74}
\begin{barticle}[mr]
\bauthor{\bsnm{Wolfe},~\bfnm{Douglas~A.}\binits{D.~A.}}
(\byear{1974}).
\btitle{A characterization of population weighted-symmetry and related
  results}.
\bjournal{J.~Amer. Statist. Assoc.}
\bvolume{69}
\bpages{819--822}.
\bid{issn={0162-1459}, mr={0426239}}
\bptok{imsref}%
\end{barticle}
\endbibitem

\bibitem[\protect\citeauthoryear{Yanagawa}{1984}]{Yan84}
\begin{barticle}[mr]
\bauthor{\bsnm{Yanagawa},~\bfnm{Takashi}\binits{T.}}
(\byear{1984}).
\btitle{Case-control studies: Assessing the effect of a confounding factor}.
\bjournal{Biometrika}
\bvolume{71}
\bpages{191--194}.
\bid{doi={10.1093/biomet/71.1.191}, issn={0006-3444}, mr={0738341}}
\bptok{imsref}%
\end{barticle}
\endbibitem

\bibitem[\protect\citeauthoryear{Yu and Gastwirth}{2005}]{YuGas05}
\begin{barticle}[auto:STB|2011/12/02|17:21:01]
\bauthor{\bsnm{Yu},~\bfnm{B.~B.}\binits{B.~B.}} \AND
  \bauthor{\bsnm{Gastwirth},~\bfnm{J.~L.}\binits{J.~L.}}
(\byear{2005}).
\btitle{Sensitivity analysis for trend tests: Application to the risk of
  radiation exposure}.
\bjournal{Biostatistics}
\bvolume{6}
\bpages{201--209}.
\bptok{imsref}%
\end{barticle}
\endbibitem

\end{thebibliography}
\end{document}